\documentclass[journal=jacsat,manuscript=article]{achemso}

\usepackage[version=3]{mhchem} 
\usepackage{xcolor}
\usepackage{float}
\usepackage{graphicx}
\usepackage{hyperref}
\usepackage{subcaption}
\usepackage{caption}

\author{Saeid Izadshenas}
\affiliation
{Institute of Physics, Faculty of Physics, Astronomy and Informatics, Nicolaus Copernicus University in Toru\'{n}, Gagarina 11, 87-100 Toru\'{n}, Poland}
\email{s.izadshenas@doktorant.umk.pl}
\author{Karolina S\l{}owik}
\affiliation
{Institute of Physics, Faculty of Physics, Astronomy and Informatics, Nicolaus Copernicus University in Toru\'{n}, Gagarina 11, 87-100 Toru\'{n}, Poland}

\title{Molecular saturation determines distinct plasmonic enhancement scenarios for two-photon absorption signal}

\abbreviations{IR,NMR,UV}

\begin{document}







\begin{abstract}
  Two-photon absorption in molecules, of significance for high-resolution imaging applications, is typically characterised with low cross sections. To enhance the TPA signal, one effective approach exploits plasmonic enhancement. For this method to be efficient, it must meet several criteria, including broadband operational capability and a high fluorescence rate to ensure effective signal detection. In this context, we introduce a novel plus-shaped silver nanostructure designed to exploit the coupling of bright and dark plasmonic modes. This configuration considerably improves both the absorption and fluorescence of molecules across near-infrared and visible spectra. By fine-tuning the geometrical parameters of the nanostructure, we align the plasmonic resonances with the optical properties of specific TPA-active dyes, i.e., ATTO 700, Rhodamine 6G, and ATTO 610. 
  
  The expected TPA signal enhancement is evaluated using classical estimations based on the assumption of independent enhancement of absorption and fluorescence. These results are then compared with outcomes obtained in a quantum-mechanical approach to evaluate the stationary photon emission rate. Our findings reveal the important role of molecular saturation determining the regimes where either absorption or fluorescence enhancement leads to an improved TPA signal intensity, considerably below the classical predictions. 

  The proposed nanostructure design not only addresses these findings, but also might serve for their experimental verification, allowing for active polarization tuning of the plasmonic response targeting the absorption, fluorescence, or both. The insight into quantum-mechanical mechanisms of plasmonic signal enhancement provided in our work is a step forward in the more effective control of light-matter interactions at the nanoscale. 
\end{abstract}

\section{Introduction}\label{sec:intro}

Two-photon absorption (TPA) is a nonlinear optical phenomenon in which two photons from a pump laser, typically within the near-infrared (NIR) range, excite a molecule from its ground to an excited state, subsequently producing visible fluorescence. The key advantages of this nonlinear process include spatial precision due to high absorption probability \cite{Garcia‐Lechuga2014Calculation} with minimized photodamage \cite{Yi2014Two-photon}, and deeper penetration \cite{Pascal2021Near-infrared}. However, the TPA process typically requires a high-intensity illumination to overcome its low cross-sections \cite{Pascal2021Near-infrared}.

To enhance the TPA signal, several methods can be employed, including material engineering with plasmonic and dielectric nanostructures \cite{Ojambati2020Efficient,yang2015nonlinear}, molecular selection \cite{Lee2001Two-photon} and light source optimization \cite{Giri2022Manipulating,Dayan2004Two,lu2022two}. These approaches focus on improving the interaction between light and matter at the molecular level to maximize TPA efficiency.

Plasmonic nanostructures can be used to enhance TPA due to their ability to concentrate electromagnetic fields at the nanoscale. This enhancement arises from localized surface plasmon resonances (LSPRs), which occur when the free electrons in the metal collectively oscillate in response to an external light source \cite{Giannini2011Plasmonic,akselrod2014probing}. This phenomenon results in a dramatic increase in the local electric field intensity, thereby boosting the efficiency of TPA by molecules subject to enhanced field \cite{Feng2015Distance-Dependent}.

To enhance both the absorption and fluorescence of molecules in TPA, nanostructures may be engineered supporting plasmonic resonances both around the absorption and the fluorescence wavelengths. One approach to achieving two distinct plasmon resonances is through plasmon-induced transparency. This method involves spatial symmetry breaking \cite{chen2012plasmon,izadshenas2018tunable}, altering incident angle \cite{song2014dynamically}, and polarization \cite{luo2021dynamical} to couple bright and dark modes and produce two split resonances. Plasmonic modes are effective in enhancing light absorption in molecules due to their strong coupling with incident light that allows to achieve high electromagnetic field intensity at the location of the molecules \cite{Chu2009Probing}. On the other hand, they can enhance fluorescence and radiate the signal towards predefined directions in the far field, facilitating its detection.
Bimodal nanostructures can simultaneously support both absorption and fluorescence. This dual-mode approach exploits the strengths of each mode to maximize overall TPA efficiency, offering a powerful strategy for optimizing light-matter interactions at the nanoscale.


In this article, we propose a plasmonic nanostructure capable of selective enhancement of absorption and fluorescence of molecules. We demonstrate the tuning capabilities of the nanostructure that can be adjusted by design to match the optical properties of several TPA active dyes, e.g., ATTO 700,
Rhodamine 6G, and ATTO 610. We estimate the signal enhancement for TPA using the classical approach, commonly applied in the community, and compare it with the quantum-mechanical predictions. We identify molecular saturation as the key factor that determines the regimes in which either the absorption- or fluorescence-enhancement efficiently occurs, and which, in consequence, limits the overall enhancement. 


\section{Semiclassical description of TPA}\label{sec:theory}

\begin{figure*}[t]
   \centering
   \begin{minipage}{0.43\textwidth}
       \centering
       \caption*{(a)}
       \includegraphics[width=1\linewidth]{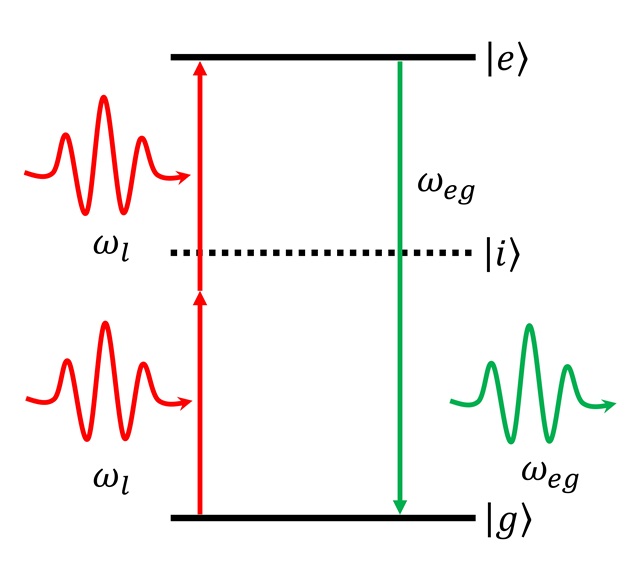}
   \end{minipage}\hfill
   \begin{minipage}{0.57\textwidth}
       \centering
       \caption*{(b)}
       \includegraphics[width=0.775\linewidth]{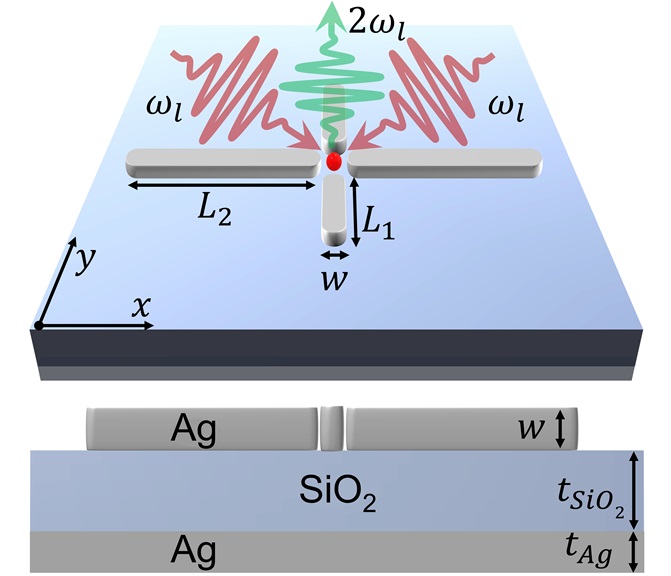}
   \end{minipage}
   \caption{Diagram (a) illustrates the TPA process where two NIR photons, absorbed by a fluorescent molecule, excite it to the excited state $|e \rangle$. This is followed by the emission of visible fluorescence as the molecule relaxes back to its ground state. (b) Illustration of a plus-shaped silver nanostructure on a $SiO_{2}$ substrate, with a red point in the gap representing a molecule. Two NIR photons (red arrows) excite the molecule, and a single fluorescence photon is emitted (green arrow).}
   \label{fig:1}
\end{figure*}

A TPA process involves a simultaneous absorption of two pump photons whose combined energy is equivalent to the transition energy necessary to excite the dye molecule from its ground to excited state. As a result, the molecule may emit a fluorescent photon, as depicted in Fig.~\ref{fig:1}(a). Since the absorption occurs through a virtual state, the probability of TPA is maximized when two photons simultaneously interact with the molecule. One technique to improve the likelihood of their simultaneous arrival at the molecular position is exploiting entangled pairs \cite{tabakaev2021energy,Giri2022Manipulating}. Another, that we investigate here, involves plasmonic nanostructures to localize light in space rather than time \cite{lu2022two}.
We begin with a consideration of a semiclassical model of a molecule positioned near a nanostructure subject to classical laser light, and indicate the field-enhancement and Purcell emission-enhancement mechanisms through which the plasmonic nanostructure can influence its stationary state.

We consider a two-level model of the molecule, with the ground and excited states $|g\rangle$ and $|e\rangle$, respectively with energies $\omega_g$ and $\omega_e$. The intermediate states, denoted as $|i\rangle$ in the scheme in Fig.~\ref{fig:1}(a), can be integrated out as derived in \textit{Supplementary Material
: Effective two-level description}.
We allow the molecule to be positioned near the nanostructure [Fig.~\ref{fig:1}(b)] that scatters the incoming plane wave of amplitude $\mathbf{E}_0(\omega_l)$, and gives rise to the field distribution $\mathbf{E}(\omega_l,\mathbf{r})$.
The resulting Hamiltonian of the molecule near the nanostructure subject to a classical continuous-wave illumination at frequency $\omega_l$, can be written in the form
\begin{equation}\label{eq:hamiltonian}
    H = \frac{1}{2}\hbar\omega_{eg}\sigma_z + \hbar \left(\Omega^{(2)}_\mathrm{NP}e^{-2i\omega_l t}\sigma_++{\Omega^{(2)}}^\star_\mathrm{NP}e^{2i\omega_l t}\sigma_-\right),
\end{equation}
where $\sigma_-=|e\rangle\langle g|$ and $\sigma_-=\sigma_+^\dagger$ are Pauli flip operators describing molecular transitions between the ground and excited states, $\sigma_z=|e\rangle\langle e|-|g\rangle\langle g|$ is the population inversion operator, $\omega_{eg}=\omega_e-\omega_g$ is the transition frequency from the ground state to excited state, and $\Omega^{(2)}_\mathrm{NP}$ is the two-photon coupling strength with the molecule in presence of the nanostructure.
In \textit{Supplementary Material
: Effective two-level description}, this quantity is derived in the form
\begin{equation}
    \Omega^{(2)}_\mathrm{NP}=\sum_i\underbrace{\frac{E(\omega_l,\mathbf{r}_m)}{E_{0}(\omega_l)}
    \frac{E(\omega_l,\mathbf{r}_m)}{E_{0}(\omega_l)}}_\mathrm{field\ enhancement\ factors}
    \Omega^{(2)}_i
    \label{eq:Rabi}
\end{equation}
being a sum of contributions arising in the presence of virtual states $|i\rangle$ of energy $\hbar\omega_i$ being considerably detuned from single-photon resonances: $|2\omega_l-\omega_{eg}| \ll |\omega_l-\omega_{ig}|, |\omega_l-\omega_{ei}| \ll \omega_l$. Each contribution is a product of the field enhancement factors and the free-space effective two-photon coupling $\Omega^{(2)}_i$.
The field enhancement factors are ratios of the electric field of amplitude $E(\omega_l,\mathbf{r})=|\mathbf{E}(\omega_l,\mathbf{r}_m)|$ evaluated at the molecular position $\mathbf{r}_m$, and the plane wave amplitude $E_0(\omega_l) = |\mathbf{E}_0(\omega_l)|$.
The effective two-photon coupling takes the form $\Omega^{(2)}_i = \frac{\Omega^{(1)}_{ig}\Omega^{(1)}_{ei}}{\omega_l-\omega_{ig}}$. In this expression, $\Omega^{(1)}_{ei},\Omega^{(1)}_{ig}$ are the coupling strengths of single-photon molecular transitions involving the virtual states, $\omega_l-\omega_{ig}$ is the detuning between the laser frequency and the transition frequency between the ground and virtual intermediate levels $\omega_{ig}=\omega_i-\omega_g$. Deriving Eq.~(\ref{eq:Rabi}), we have assumed all relevant molecular transition dipole moment elements to be co-oriented. 

Once the TPA excites the molecule from $|g\rangle$ to $|e\rangle$, two main scenarios can be considered for its return to the ground state: a two-photon transision through the $\sigma_-$ Hamiltonian term (\ref{eq:hamiltonian}), or an emission of a single fluorescent photon. In this section, we assume that fluorescence occurs at $\omega_{eg}$, as depicted in Fig.~\ref{fig:1}(a) with the green arrow. In practice, fluorescence serves as a background-free evidence of TPA in experimental scenarios \cite{tabakaev2021energy}. To account for fluorescence, we turn to a density matrix description of the molecular state. Its stationary form can be found from the Gorini - Kossakowski - Lindblad - Sudarshan equation, written here for the steady state $\rho$ \cite{gorini1976completely, lindblad1976generators}
\begin{equation}
    \frac{i}{\hbar}\left[H,\rho\right]-\mathcal{L}_\gamma(\rho)=0.
\end{equation}
where $\mathcal{L}_\gamma(\rho) = \gamma\left( \sigma_-\rho\sigma_+ -\frac{1}{2}\{ \sigma_+\sigma_-,\rho\} \right)$ is the Lindblad operator describing spontaneous emission from $|e\rangle$ to $|g\rangle$ with the rate $\gamma$. Near the nanostructure, this rate can be enhanced \textit{via} the Purcell effect\cite{novotny2011antennas}
\begin{equation}
    \gamma = \frac{P(\omega_{eg},\mathbf{r}_m)}{P_{0}(\omega_{eg})}\gamma_0,
    \label{eq:Purcell}
\end{equation}
where $\gamma_0$ is the free-space emission rate given by the Weisskopf-Wigner formula \cite{scully1997quantum}. The ratio $\frac{P(\omega,\mathbf{r})}{P_{0}(\omega)}$ is the Purcell enhancement factor of the total (radiated and absorbed) emission power $P(\omega,\mathbf{r})$ of an electric dipole oscillating at the frequency $\omega$, positioned at $\mathbf{r}$ near the nanostructure, over the emission power $P(\omega)$ of a dipole oscillating at the same frequency in free space. In Eq.~(\ref{eq:Purcell}), the dipole is located at the molecular position $\mathbf{r}_m$ and has the frequency $\omega_{eg}$ so that it models a dipolar transition $|e\rangle \leftrightarrow |g\rangle$.

Before we continue to discuss the plasmonic impact on the TPA in this semiclassical framework, we introduce the nanostructure engineered to support the sequence of TPA and fluorescent emission.

\section{Plasmonic nanostructure}
Figure.~\ref{fig:1}(b) illustrates a plus-shaped silver nanostructure composed of four nanobars separated by a gap, situated on an $SiO_{2}$ substrate. The silver film located on the bottom of the nanostructure reflects all incident light and enhances optical response of the nanobars. The red point represents a dye molecule located in the gap between the nanobars. The incident electric field is polarized in the $xy$ plane, while light propagates in the $z$ direction.
The nanostructure geometry is designed to obtain two split resonances in near-infrared and visible regimes: Different nanobar lengths are chosen along the $x$ and $y$ directions, respectively $L_2 = 135\ \mathrm{nm}$ and $L_1 = 60\ \mathrm{nm}$. All nanobars have the same widths and heights $w = 25\ \mathrm{nm}$. 
The glass spacer thickness is $t_{SiO_2} = 50\ \mathrm{nm}$. The overall response of the nanostructure weakly depends on the bottom silver film thickness. A perfectly matched layer is assumed as domain around the plasmonic nanostructure. The metal film's length and width are $P_x = 400\ \mathrm{nm}$ and $P_y = 400\ \mathrm{nm}$, and the thickness is $t_{Ag} = 60\ \mathrm{nm}$. 

\begin{figure}[h!]
    \begin{minipage}{0.33\columnwidth}
        \caption*{(a)}
        \includegraphics[width=\columnwidth]{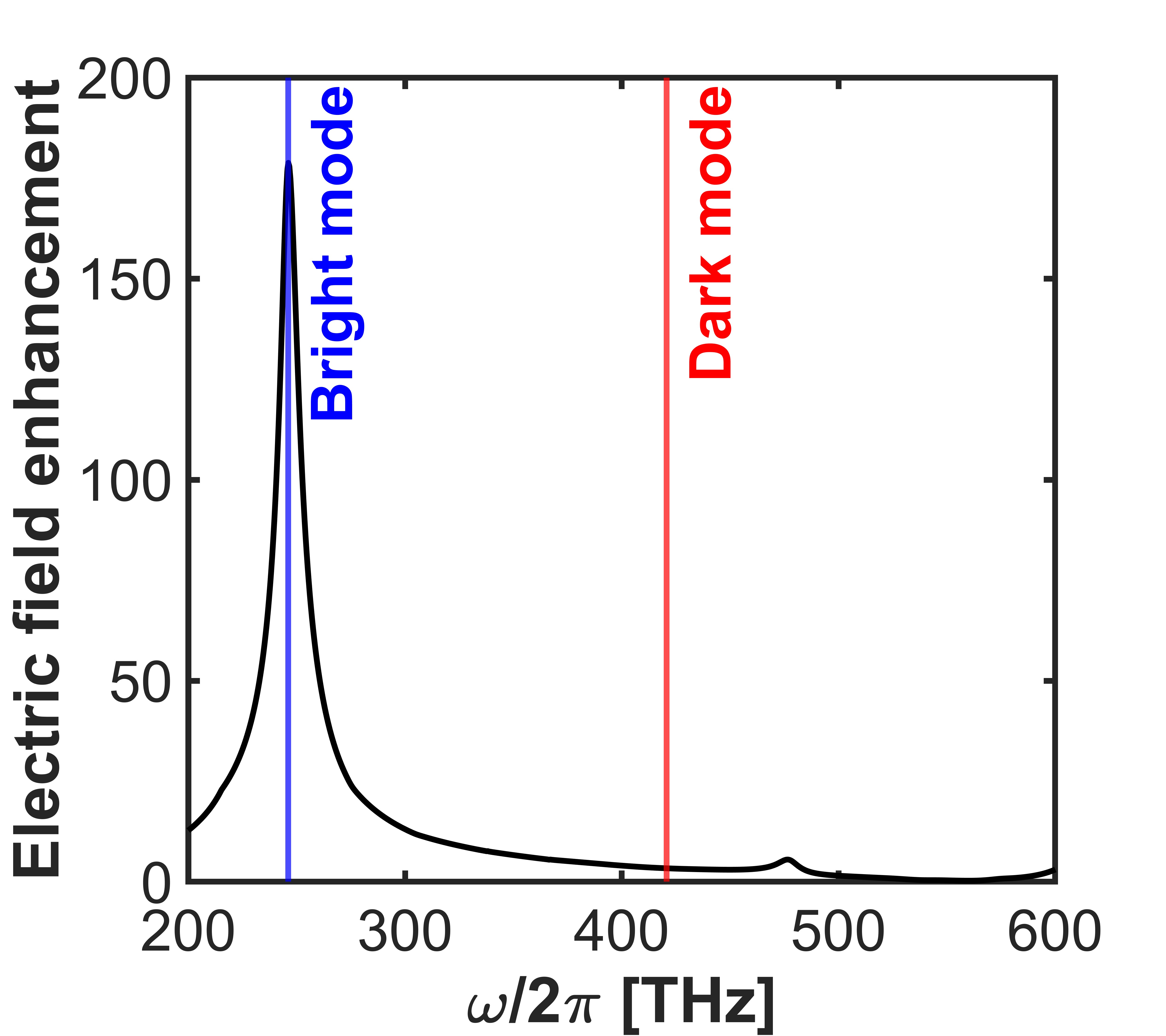}
    \end{minipage}\hfill
    \begin{minipage}{0.33\columnwidth}
        \caption*{(b)}
        \includegraphics[width=\columnwidth]{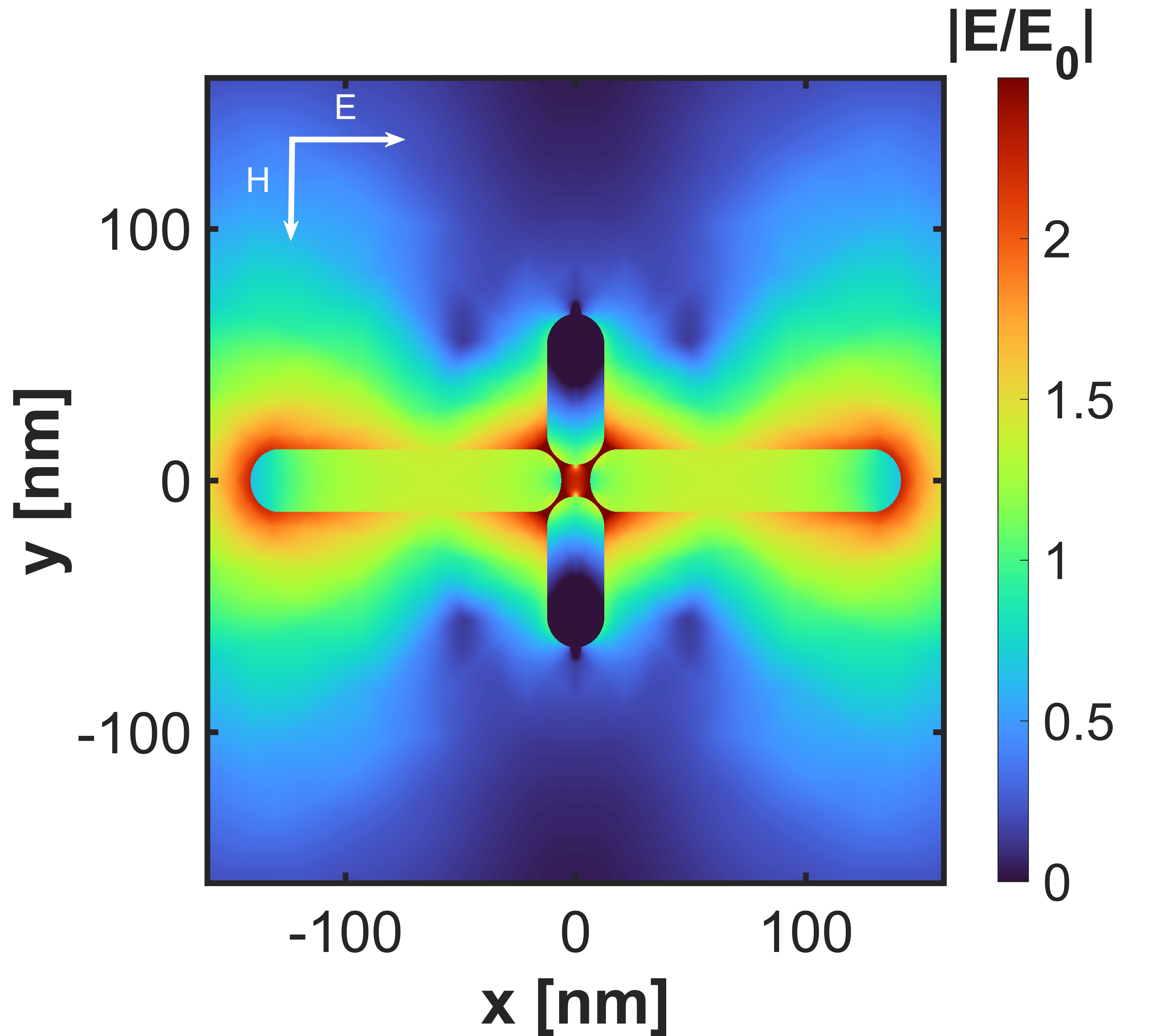}
    \end{minipage}
    \begin{minipage}{0.33\columnwidth}
        \caption*{(c)}
        \includegraphics[width=\columnwidth]{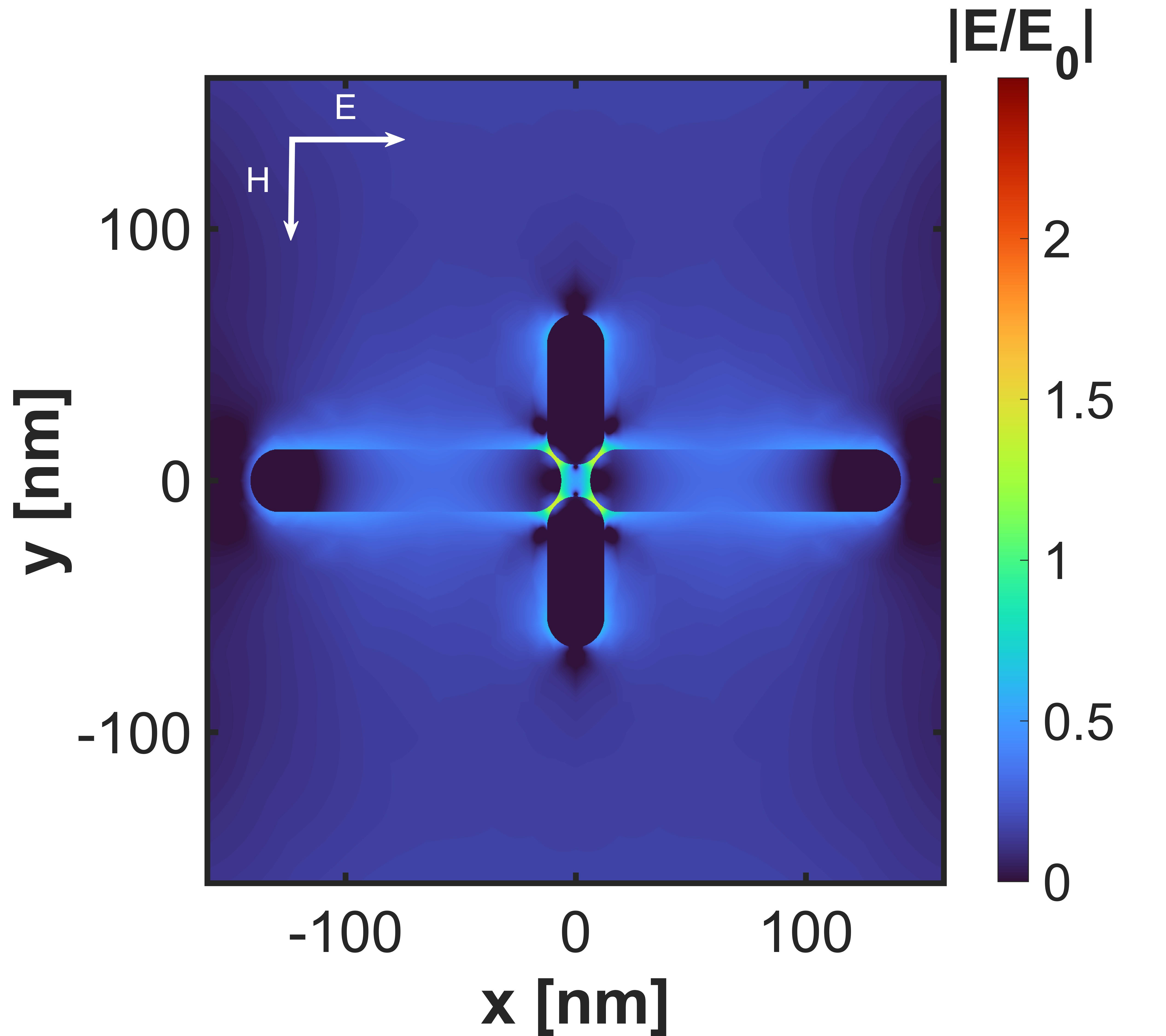}
    \end{minipage}\hfill
    \begin{minipage}{0.33\columnwidth}
        \caption*{(d)}
        \includegraphics[width=\columnwidth]{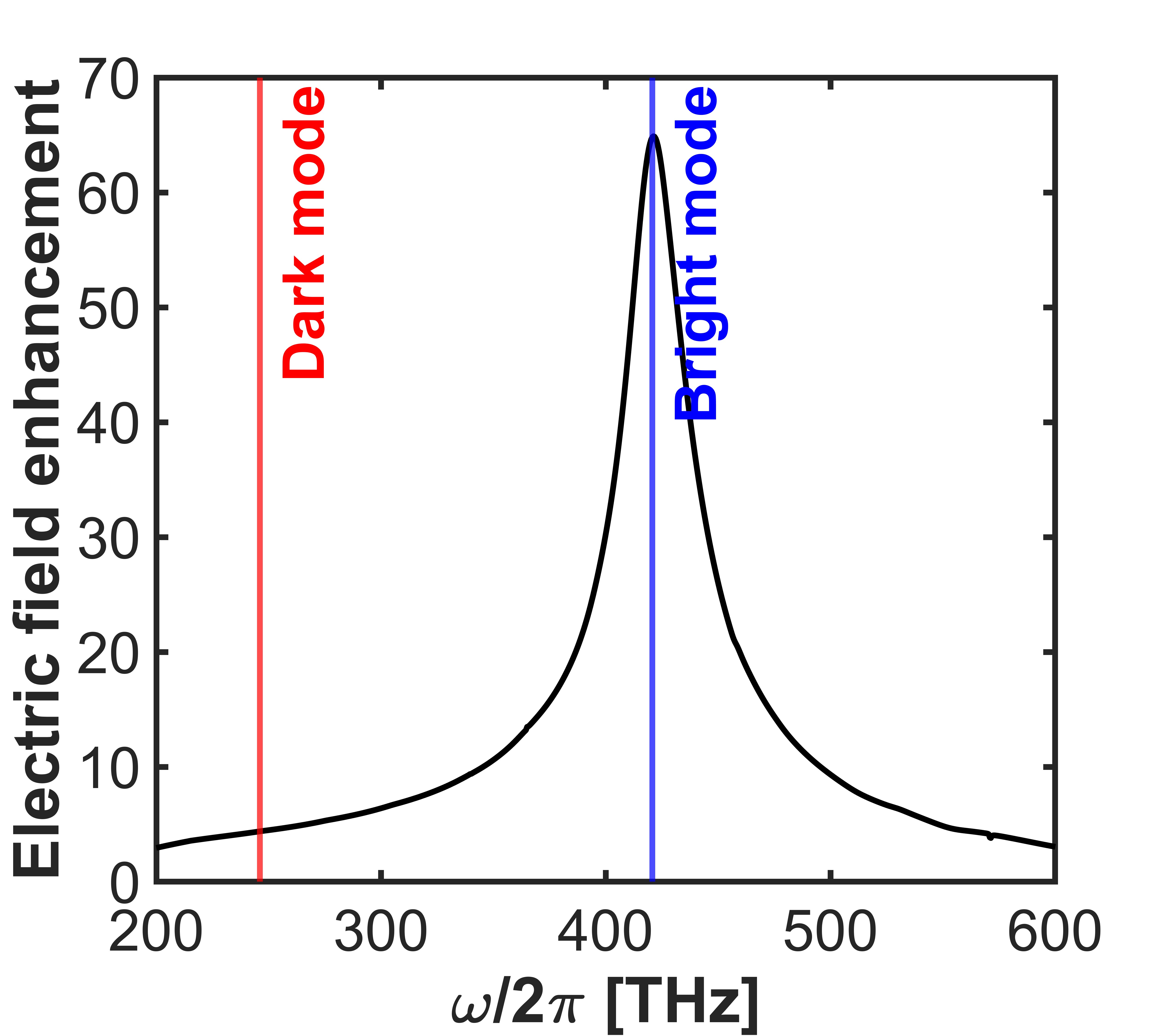}
    \end{minipage}\hfill
    \begin{minipage}{0.33\columnwidth}
        \caption*{(e)}
        \includegraphics[width=\columnwidth]{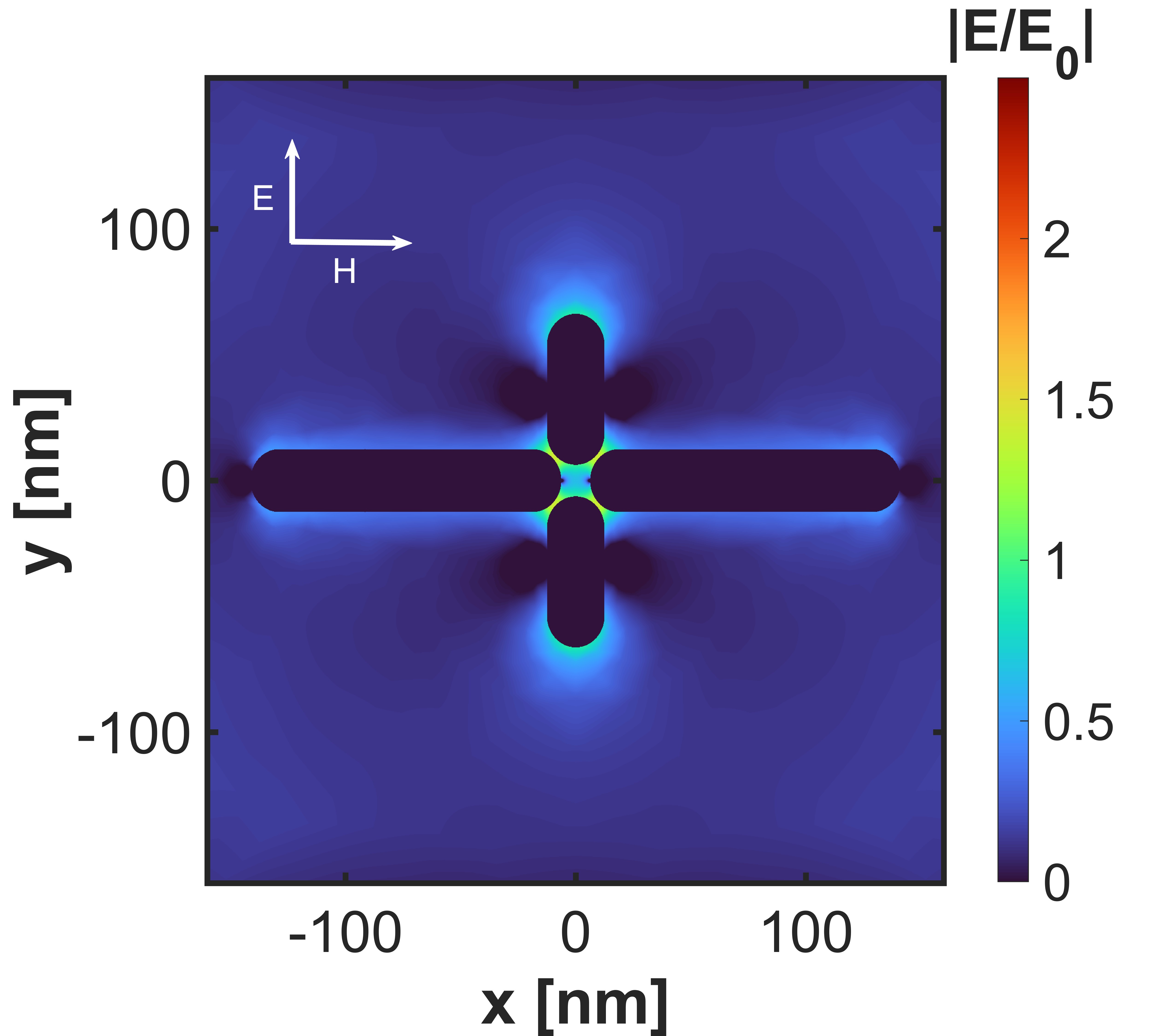}
    \end{minipage}\hfill
    \begin{minipage}{0.33\columnwidth}
        \caption*{(f)}
        \includegraphics[width=\columnwidth]{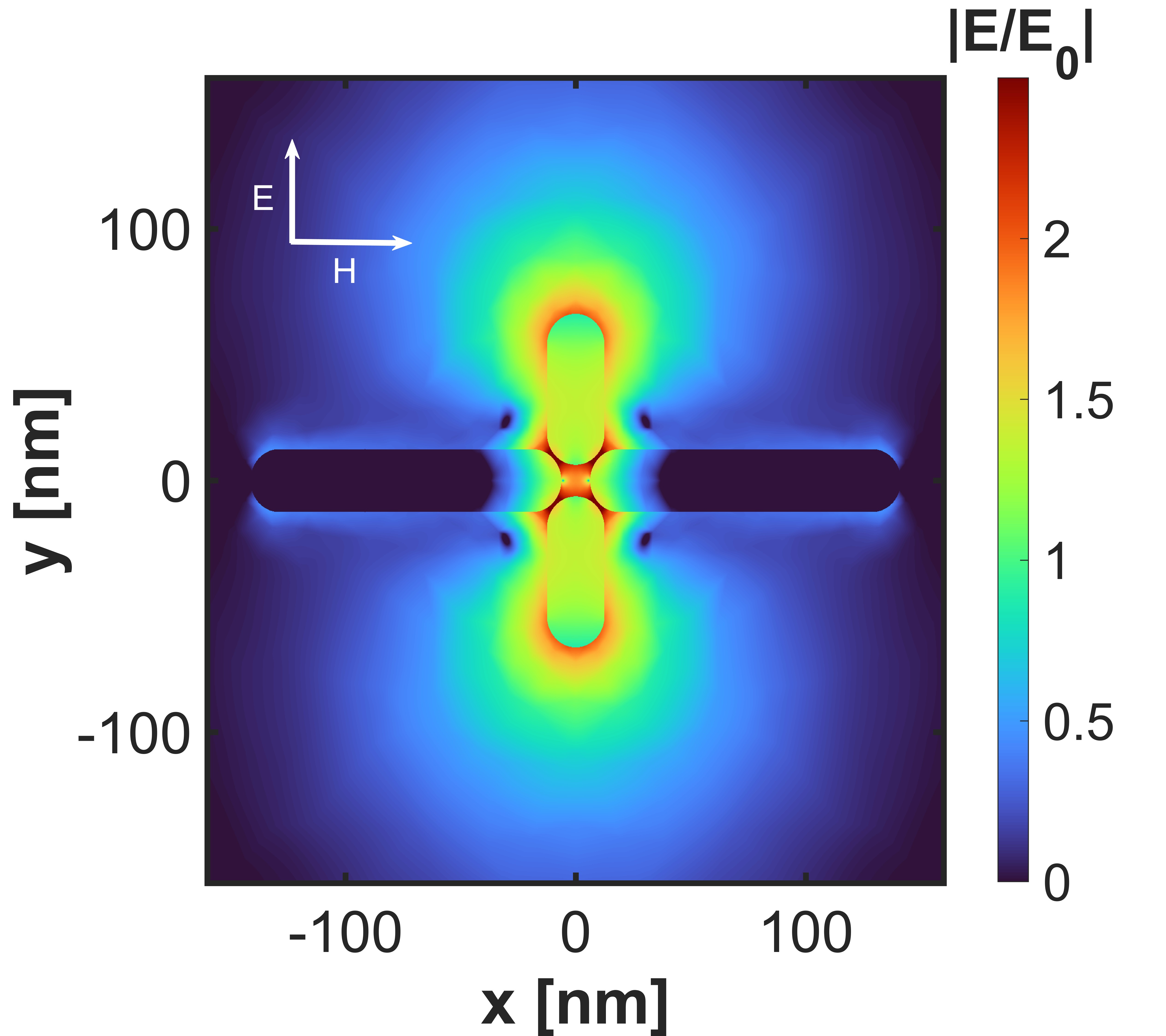}
    \end{minipage}\hfill
    \begin{minipage}{0.33\columnwidth}
        \caption*{(g)}
        \includegraphics[width=\columnwidth]{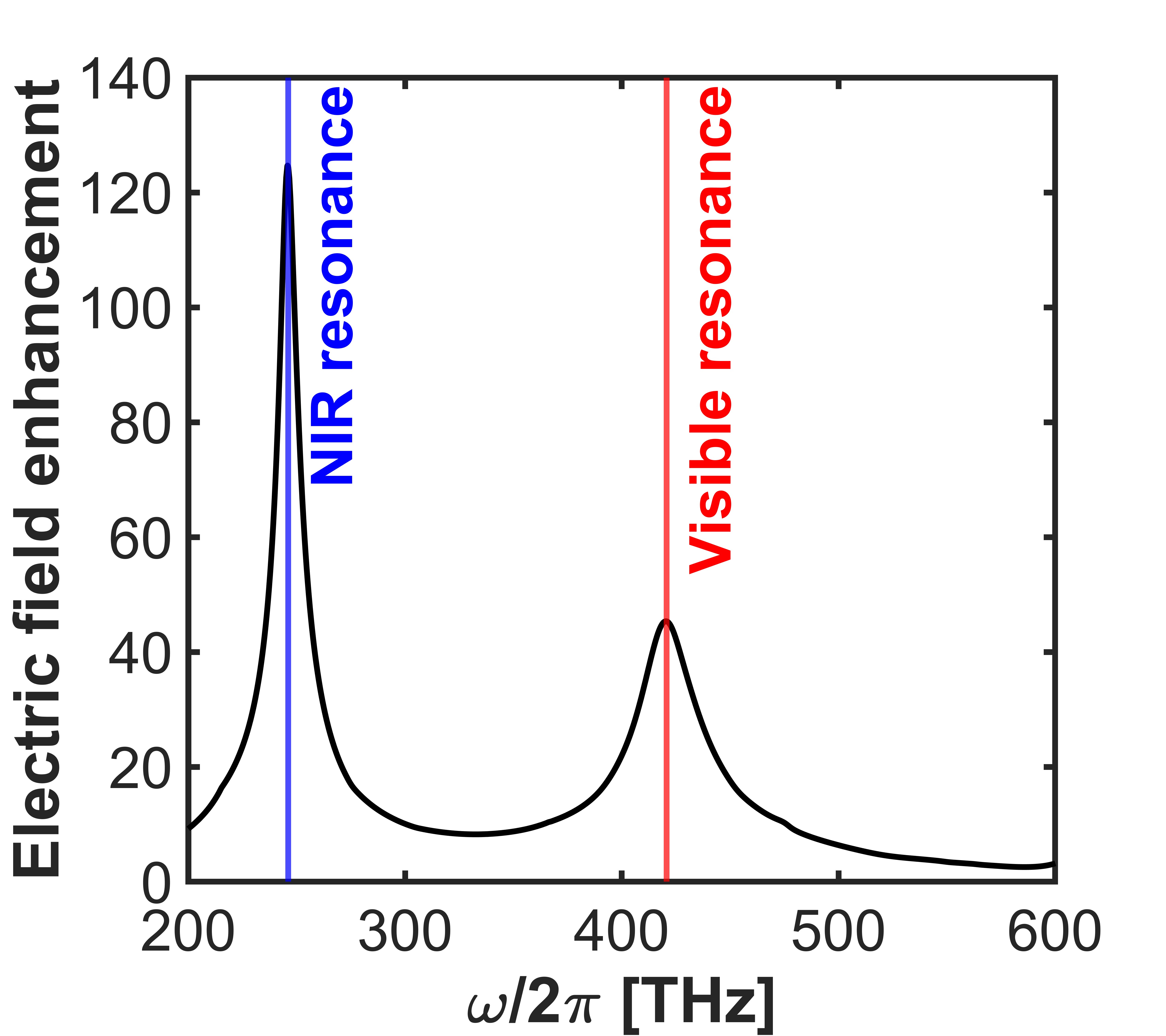}
    \end{minipage}\hfill
    \begin{minipage}{0.33\columnwidth}
        \caption*{(h)}
        \includegraphics[width=\columnwidth]{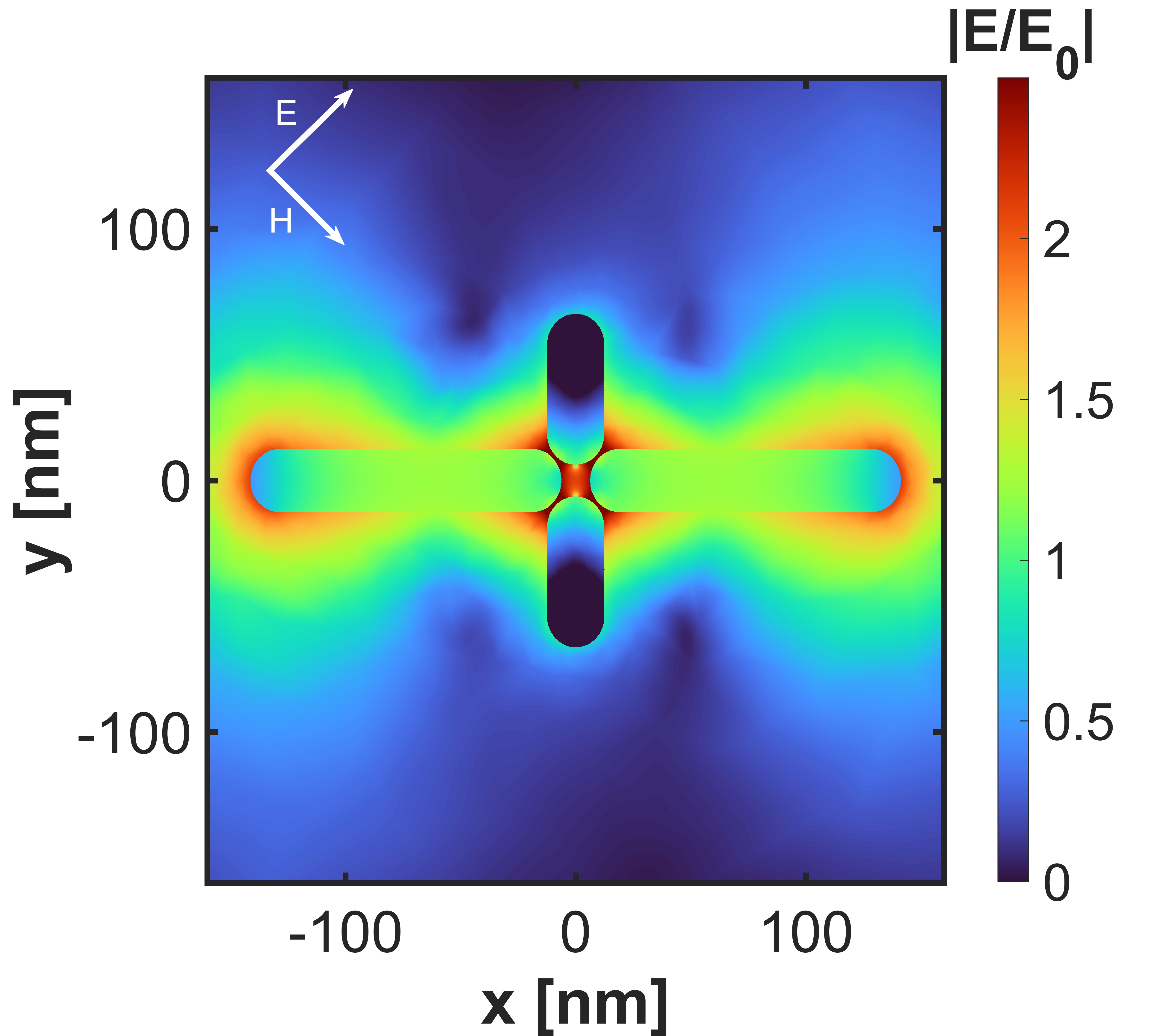}
    \end{minipage}\hfill
    \begin{minipage}{0.33\columnwidth}
        \caption*{(k)}
        \includegraphics[width=\columnwidth]{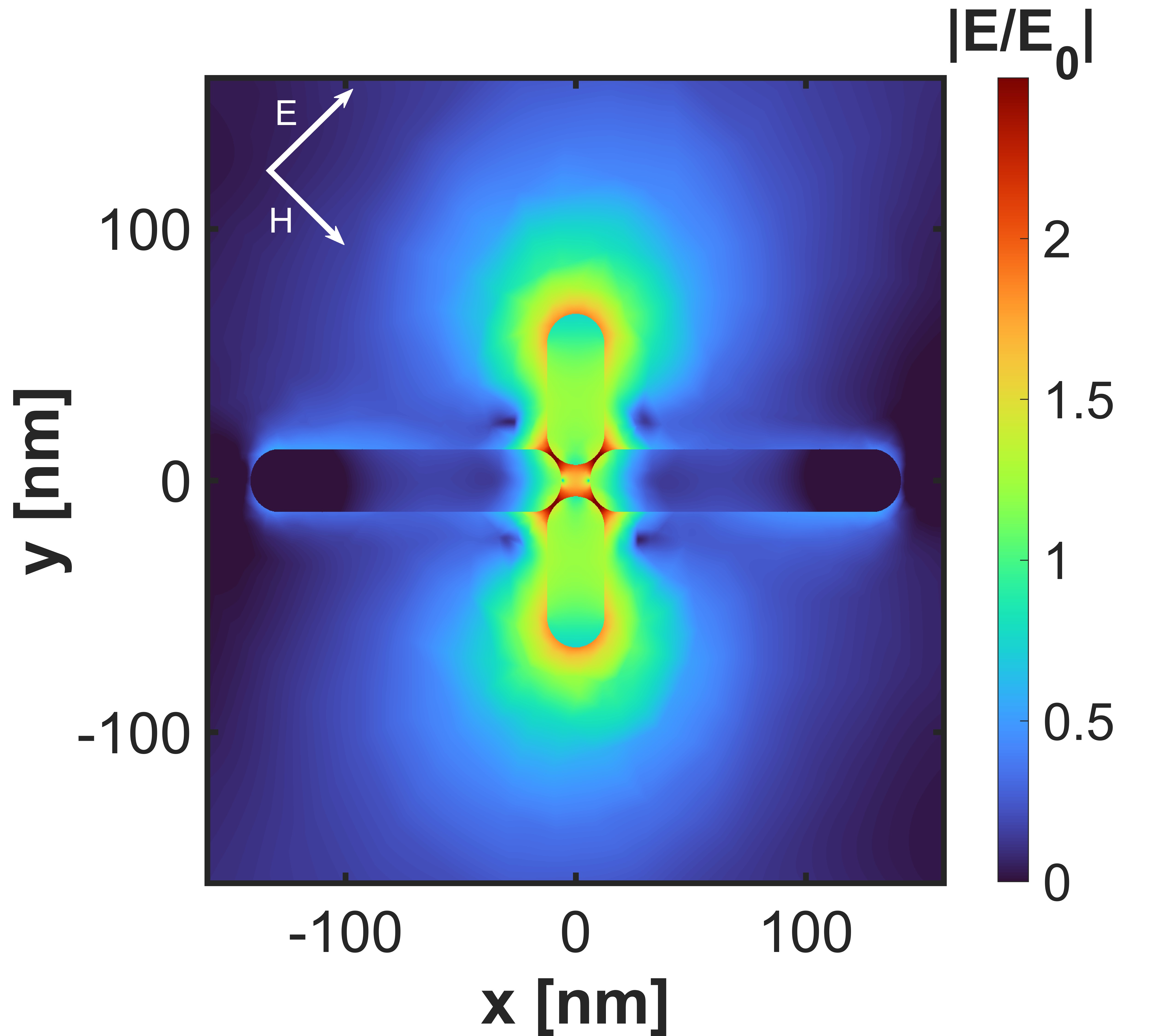}
    \end{minipage}
    \caption{Illustration of the electric field enhancement spectrum and electric field distribution for various conditions: (a) x-polarization at $\phi = 0^{\circ}$, (b) bright mode map at $\omega/2\pi = 246.1\ \mathrm{THz}$, (c) dark mode map at $\omega/2\pi = 420.6\ \mathrm{THz}$, (d) y-polarization at $\phi = 90^{\circ}$, (e) bright mode map at $\omega/2\pi = 246.1\ \mathrm{THz}$, (f) dark mode map at $\omega/2\pi = 420.6\ \mathrm{THz}$, (g) xy-polarization at $\phi = 45^{\circ}$, (h) bright mode map at $\omega/2\pi = 246.1\ \mathrm{THz}$, and (k) dark mode map at $\omega/2\pi = 420.6\ \mathrm{THz}$}
    \label{fig:E-monitor-polarization}
\end{figure}

\begin{figure}
    \centering
    \includegraphics[width=0.33\linewidth]{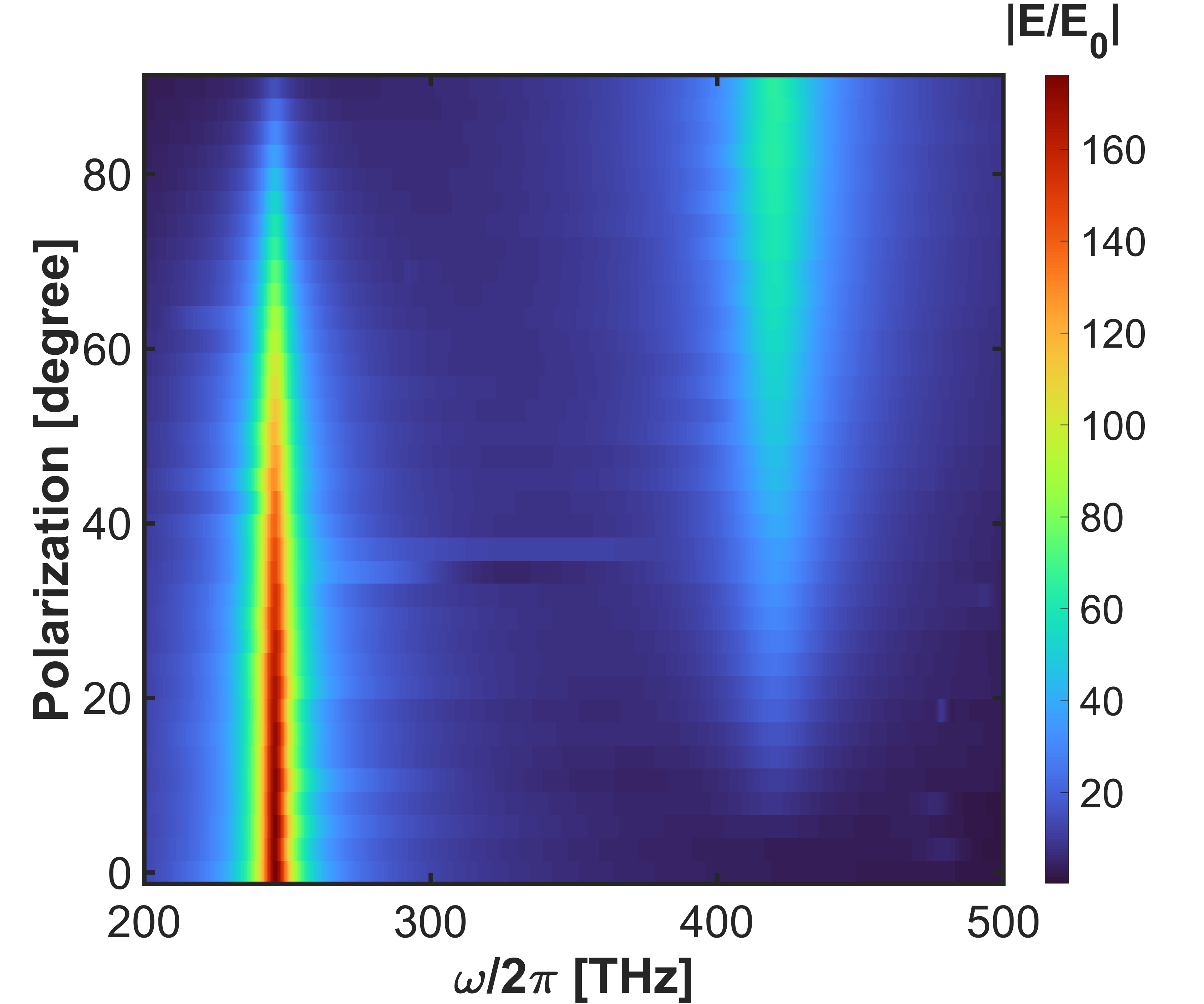}
    \caption{Electric field enhancement spectrum varying by polarization.}
    \label{fig:polarization}
\end{figure}

Using nanobars of different lengths gives rise to a pair of plasmonic resonances, which we refer to as NIR and visible  modes due to their spectral characteristics. The NIR mode is spectrally tuned to enhance absorption, while the visible one boosts fluorescence.
The polarization angle $\phi$ and illumination frequency of the incident light allow us to selectively address these resonances.
To characterize them, we investigate three different polarizations of the incident beam: $x$-polarization at $\phi = 0^{\circ}$, $y$-polarization at $\phi = 90^{\circ}$, and $xy$-polarization at $\phi = 45^{\circ}$, being the superposition of the $x$ and $y$. The NIR resonance occurs under $x$-polarized illumination, and the visible resonance is excited for the $y$-oriented polarization. By setting $xy$-polarization, we couple both NIR and visible resonances. The NIR resonance, visible resonance, and the coupled bright-dark mode in $xy$-polarization are characterized in Fig.~\ref{fig:E-monitor-polarization}.
When $x$-polarized light is applied, a single plasmonic NIR resonance at 246.1 THz is observed in the electric field enhancement spectrum $E(\omega,\mathbf{r}_m)/E_0(\omega)$ [Fig.~\ref{fig:E-monitor-polarization}(a)]. On NIR resonance, the nanostructure's scattering is dominated by longer nanobars, as it may be concluded from the bright response seen in the field enhancement distribution shown in Fig.~\ref{fig:E-monitor-polarization}(b). In Fig.~\ref{fig:E-monitor-polarization}(c), we also show the field distribution for $x$-polarized input beam at 420.6 THz, where the visible mode is located, but it turns out to be dark for this illumination scenario.
When the polarization is aligned withe the $y$ axis, a single resonance appears at 420.6 THz, i.e.,  in the visible regime [Fig.~\ref{fig:E-monitor-polarization}(d)], with a lower resonant electric field enhancement. The resonance at 246.1 THz is dark under $y$-polarized illumination [Fig.\ref{fig:E-monitor-polarization}(e)], while the one at 420.6 THz becomes bright [Fig.~\ref{fig:E-monitor-polarization}(f)] mainly due to the interaction with the shorter $y$-oriented nanobars. When the polarization is oriented at $\phi=45^{\circ}$, both visible and near-infrared resonances are excited, resulting in a coupling of the NIR and visible modes, as illustrated in the field enhancement spectrum in Fig.~\ref{fig:E-monitor-polarization}(g) and field distributions in Fig.~\ref{fig:E-monitor-polarization}(h,u), respectively for illumination frequencies of 246.1 and 420.6 THz. Thus, by adjusting the polarization, we switch the character of the plasmonic mode from dark to bright.
By sweeping the polarization angle $\phi=0^{\circ}$ to $\phi=90^{\circ}$, we observe a smooth transition between the above-discussed cases (Fig.\ref{fig:polarization}).

Having characterized the resonances in the plasmonic response, we aim to match the nanostructure's geometry to the absorption and fluorescence profile of the molecule. This can be done as the resonances can be controlled independently by two parameters:
The length of the longer nanobars affects the resonance position and the electric field enhancement $\frac{E(\omega_l,\mathbf{r}_m)}{E_{0}(\omega_l)}$ in the near-infrared regime, leading to the light-matter interaction strength enhancement according to Eq.~(\ref{eq:Rabi}).
The length of the shorter nanobars influences the resonance position and radiated power enhancement in the visible regime, boosting the Purcell enhancement $\frac{P(\omega_{eg},\mathbf{r}_m)}{P_{0}(\omega_{eg})}$ of the fluorescence rate, as stated in Eq.~(\ref{eq:Purcell}).
This independent control allows optimizing the performance of plasmonic nanostructure in the nonlinear process of TPA.
Below, we investigate the nanostructure tunability with these key parameters in detail, focusing on the field enhancement at the highest symmetry point indicated by the red dot in Fig.~\ref{fig:1}(b) for the NIR resonance, and at the power enhancement with the dipolar source positioned at the same point for the visible resonance.

\begin{figure*}[ht!]
   \centering
   \begin{minipage}{0.25\textwidth}
       \centering
       \caption*{(a)}
       \includegraphics[width=\linewidth]{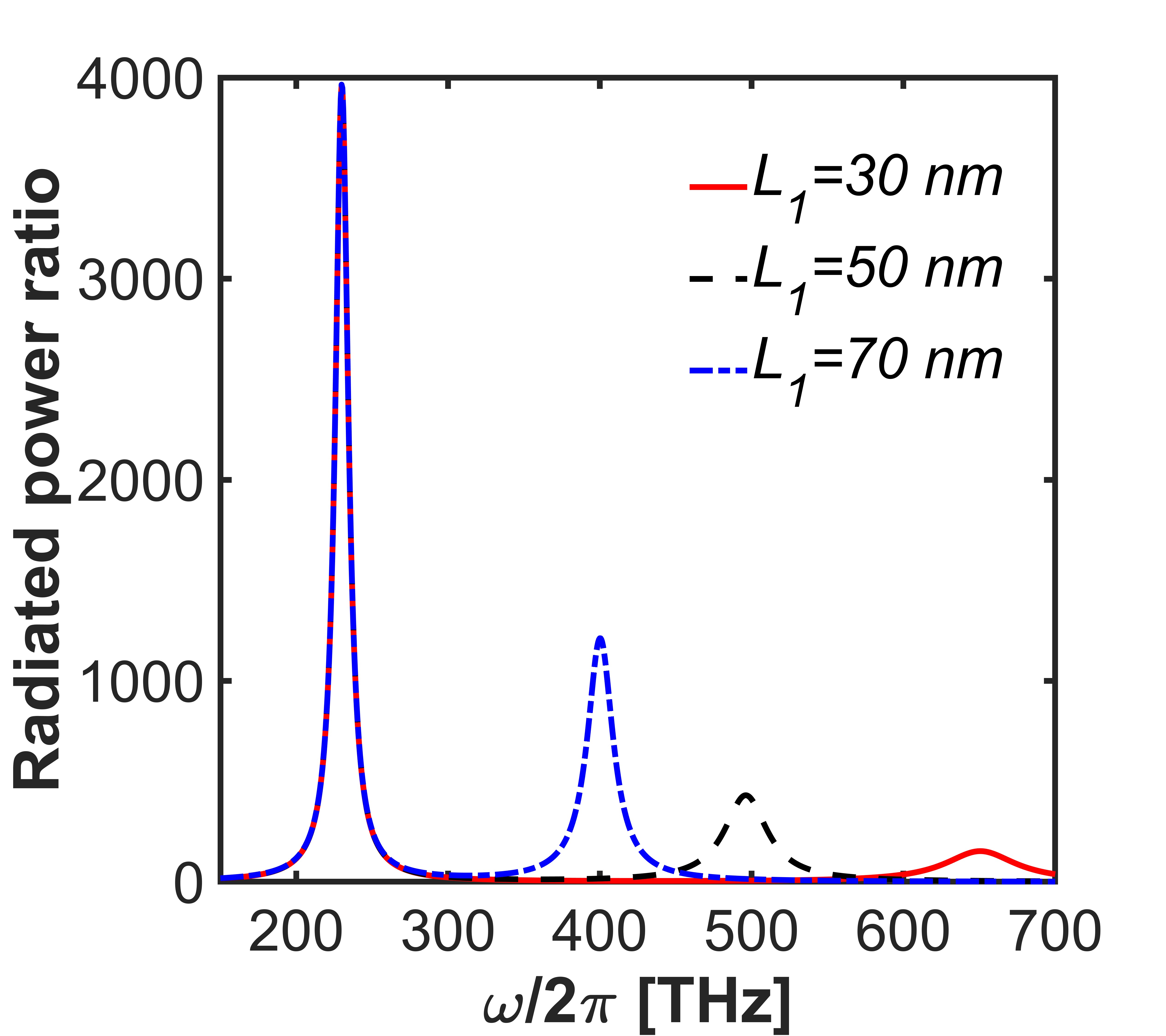} 
   \end{minipage}\hfill
   \begin{minipage}{0.25\textwidth}
       \centering
       \caption*{(b)}
       \includegraphics[width=\linewidth]{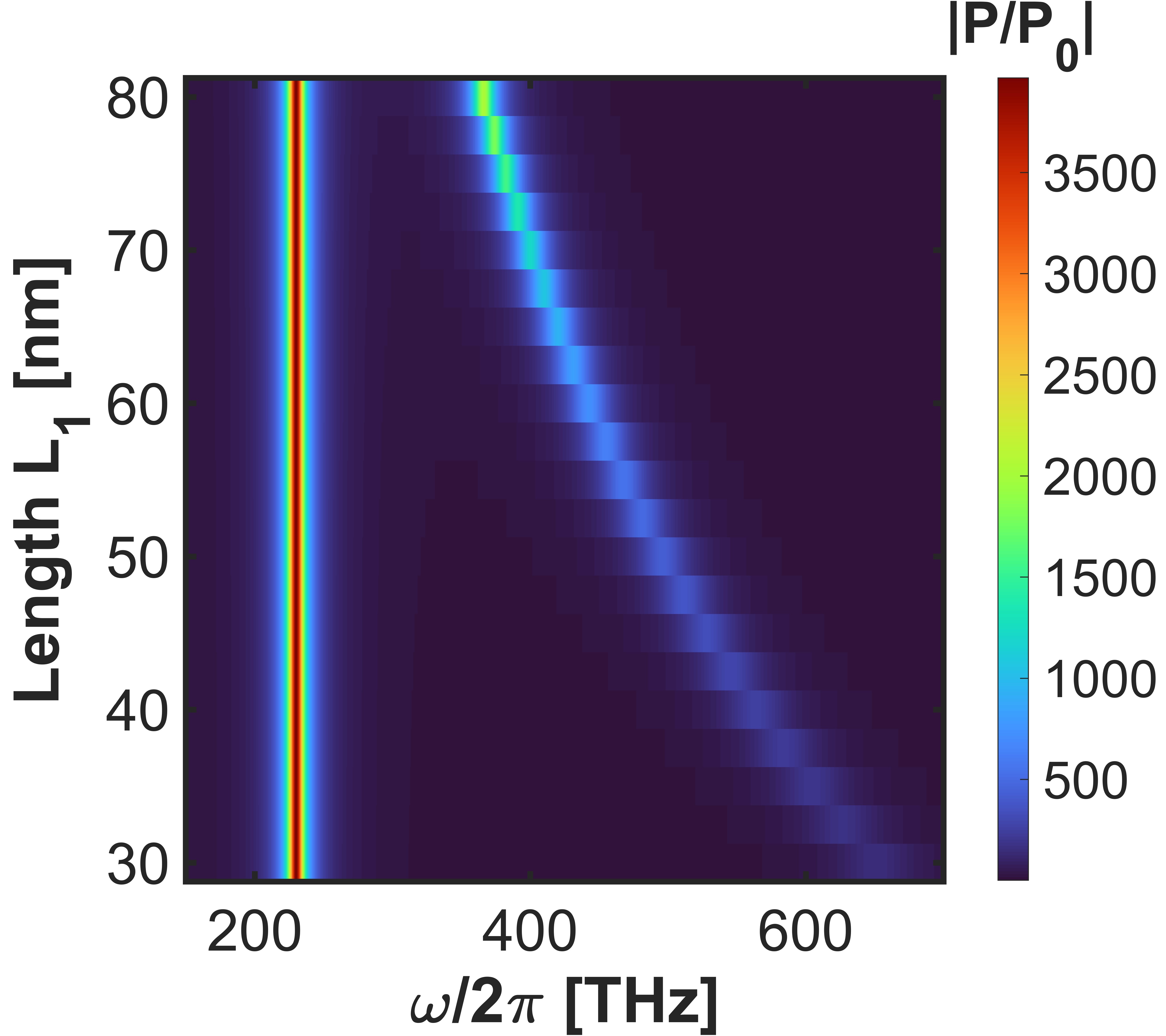} 
   \end{minipage}\hfill
   \begin{minipage}{0.25\textwidth}
       \centering
       \caption*{(c)}
       \includegraphics[width=\linewidth]{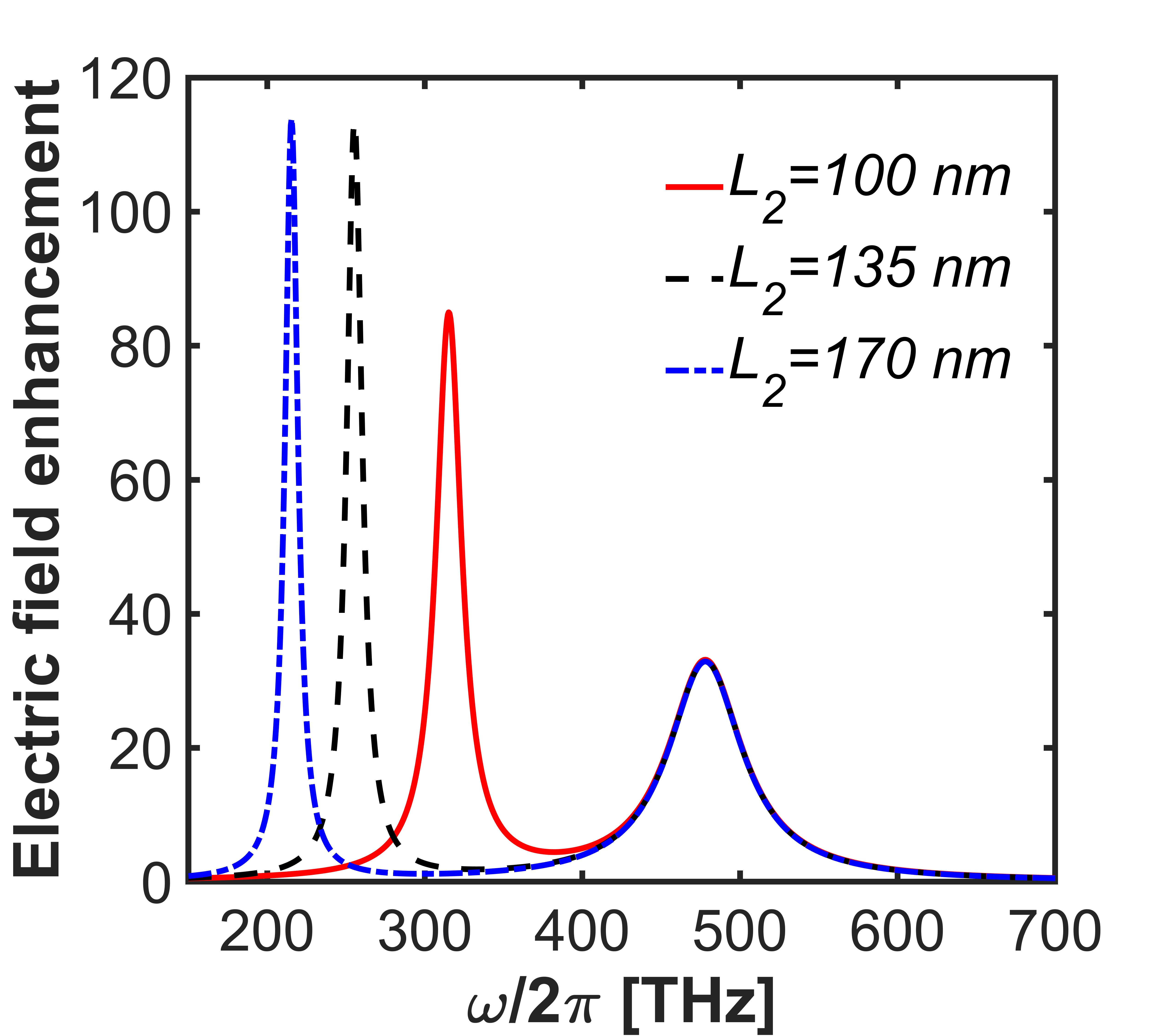} 
   \end{minipage}\hfill
   \begin{minipage}{0.25\textwidth}
       \centering
       \caption*{(d)}
       \includegraphics[width=\linewidth]{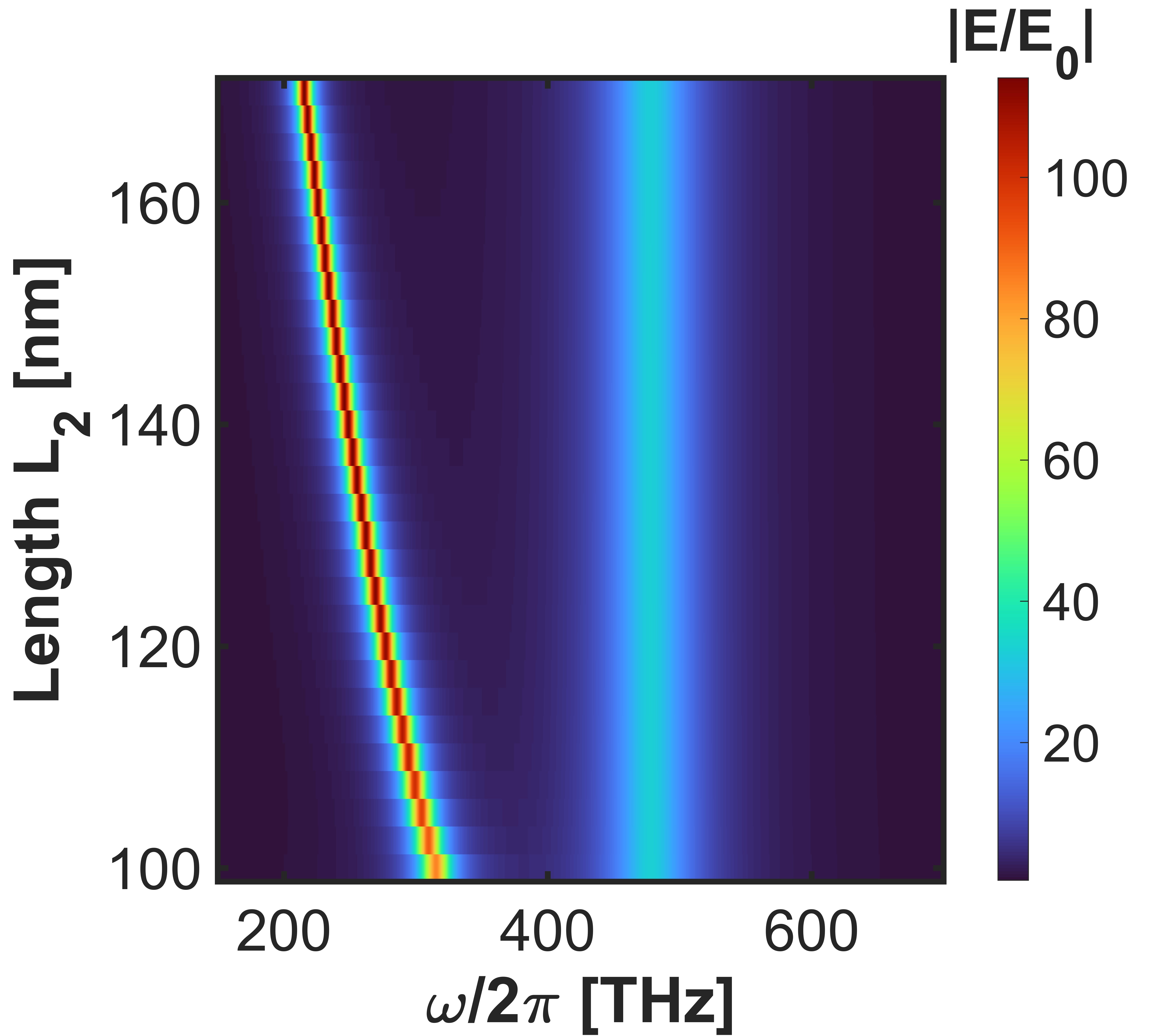} 
   \end{minipage}
   \caption{(a) Radiated power, and (b) radiated power map by varying $L_{1}$ with mirror film. (c) Electric field enhancement, and (d) electric field enhancement map by varying $L_{2}$ with mirror film.}
   \label{fig:3}
\end{figure*}

As shown in Fig.~\ref{fig:3}(a,b), increasing the size of the shorter nanobars results in a red-shift of the resonance frequency and an increase in the radiated power within the visible regime. Conversely, by keeping the dimensions of the shorter nanobars fixed and varying the size of the longer nanobars, we shift the NIR resonance while maintaining the visible resonance at a fixed spectral position, as depicted in Fig.~\ref{fig:3}(c,d). Increasing the size of the longer nanobars not only leads to a red-shift in the nanostructure's optical response but also significantly enhances the electric field amplitude. For a comparison, we provide similar results without the bottom mirror film in \textit{Supplementary Material: Plasmonic nanostructure without mirror film}. We now investigate the role of the mirror in a greater detail.

\begin{figure*}[ht!]
    \centering
    \begin{minipage}{0.45\textwidth}
        \centering
        \caption*{(a)}
        \includegraphics[width=\linewidth]{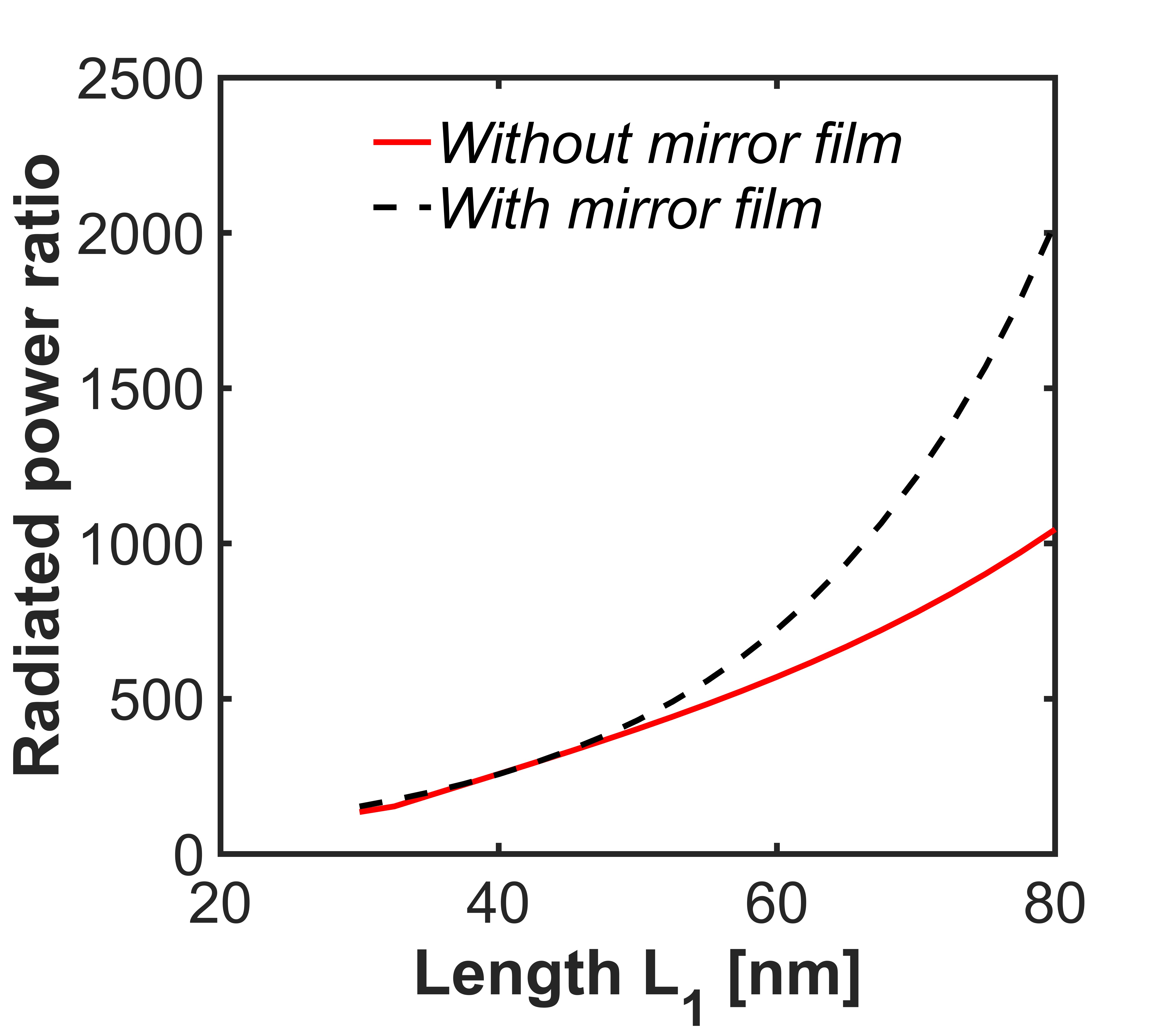}
    \end{minipage}\hfill
    \begin{minipage}{0.45\textwidth}
        \centering
        \caption*{(b)}
        \includegraphics[width=\linewidth]{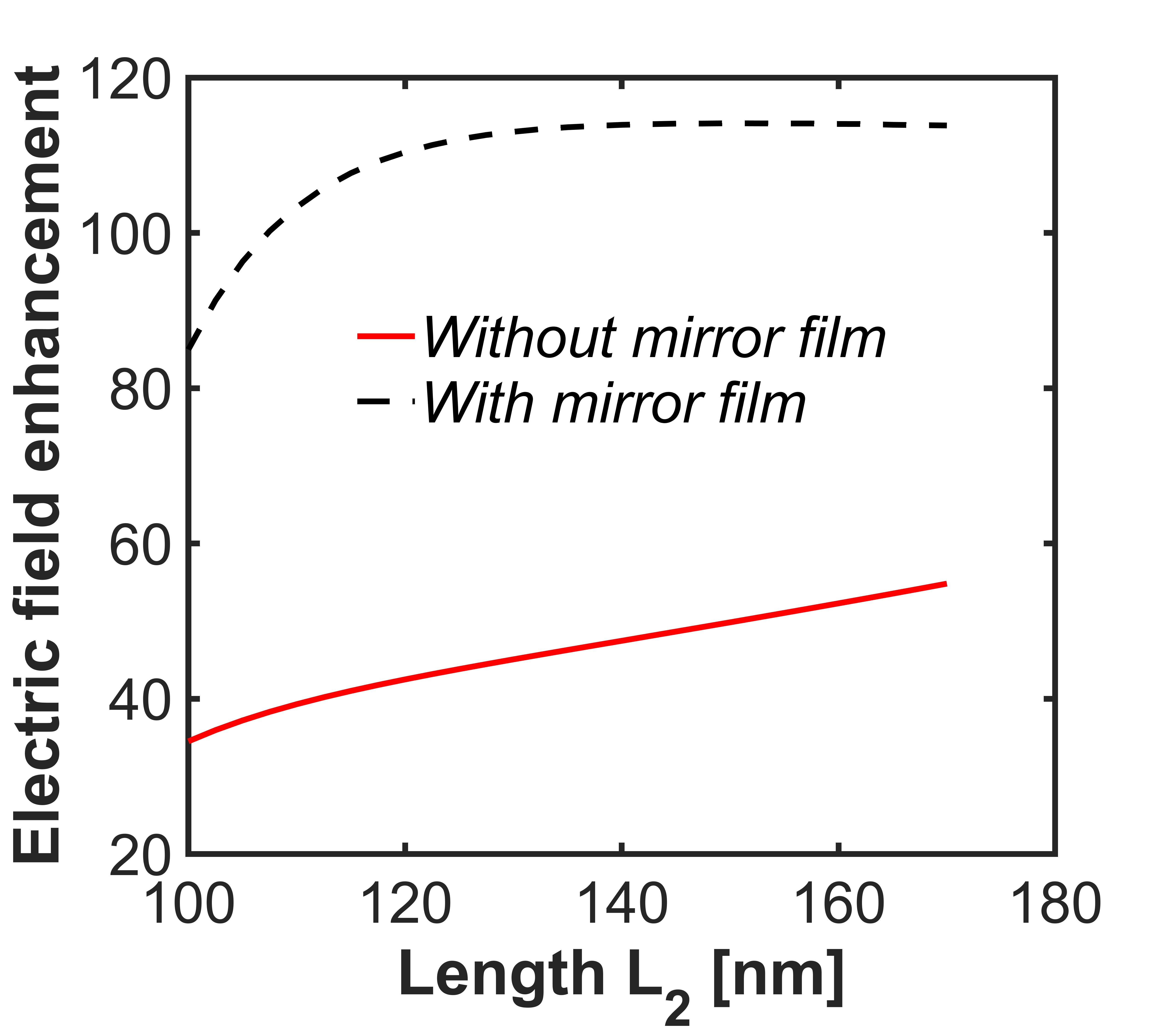}
    \end{minipage}
    \caption{(a) Radiated power ratio, (b) electric field enhancement with and without mirror film}
    \label{fig:4}
\end{figure*}

By integrating a metal mirror film on the bottom of the nanostructure, we can enhance both the radiated power and the electric field intensity without altering the geometry size of the nanostructure. This substantial enhancement results from the constructive interference of the electromagnetic fields between the metal layers. As shown in Fig.~\ref{fig:4}(a), the amplitude of the radiated power increases up to two times in the investigated range of lengths compared to the configuration without the mirror film. However, in the case of smaller nanobars, the mirror film does not significantly enhance the radiated power. This is due to the limited field confinement and weaker constructive interference in these configurations.
The electric field enhancement is even larger with the mirror film, as demonstrated in Fig.~\ref{fig:4}(b).

\begin{figure*}[ht!]
   \centering
   \begin{minipage}{0.25\textwidth}
       \centering
       \caption*{(a)}
       \includegraphics[width=\linewidth]{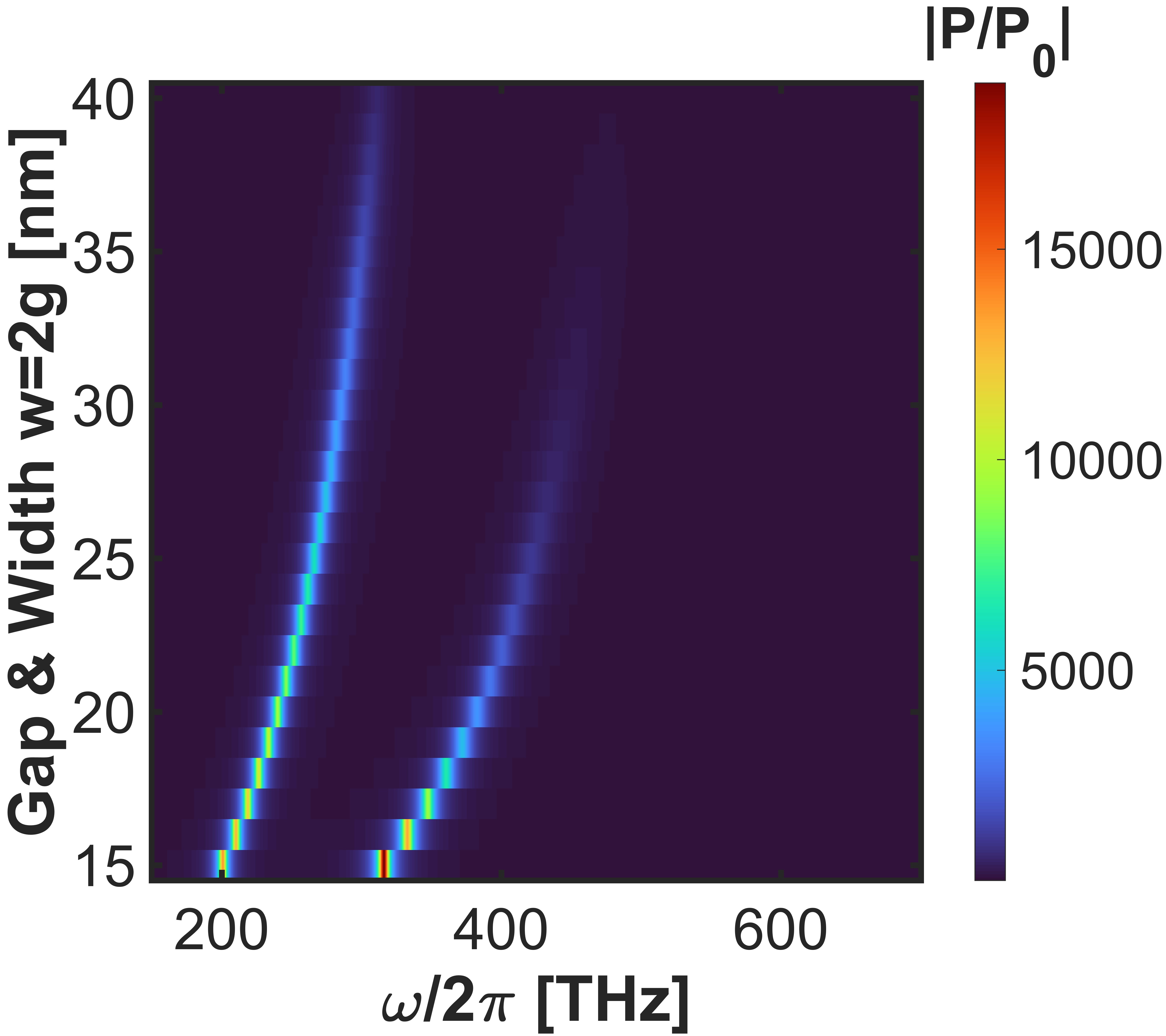} 
   \end{minipage}\hfill
   \begin{minipage}{0.25\textwidth}
       \centering
       \caption*{(b)}
       \includegraphics[width=\linewidth]{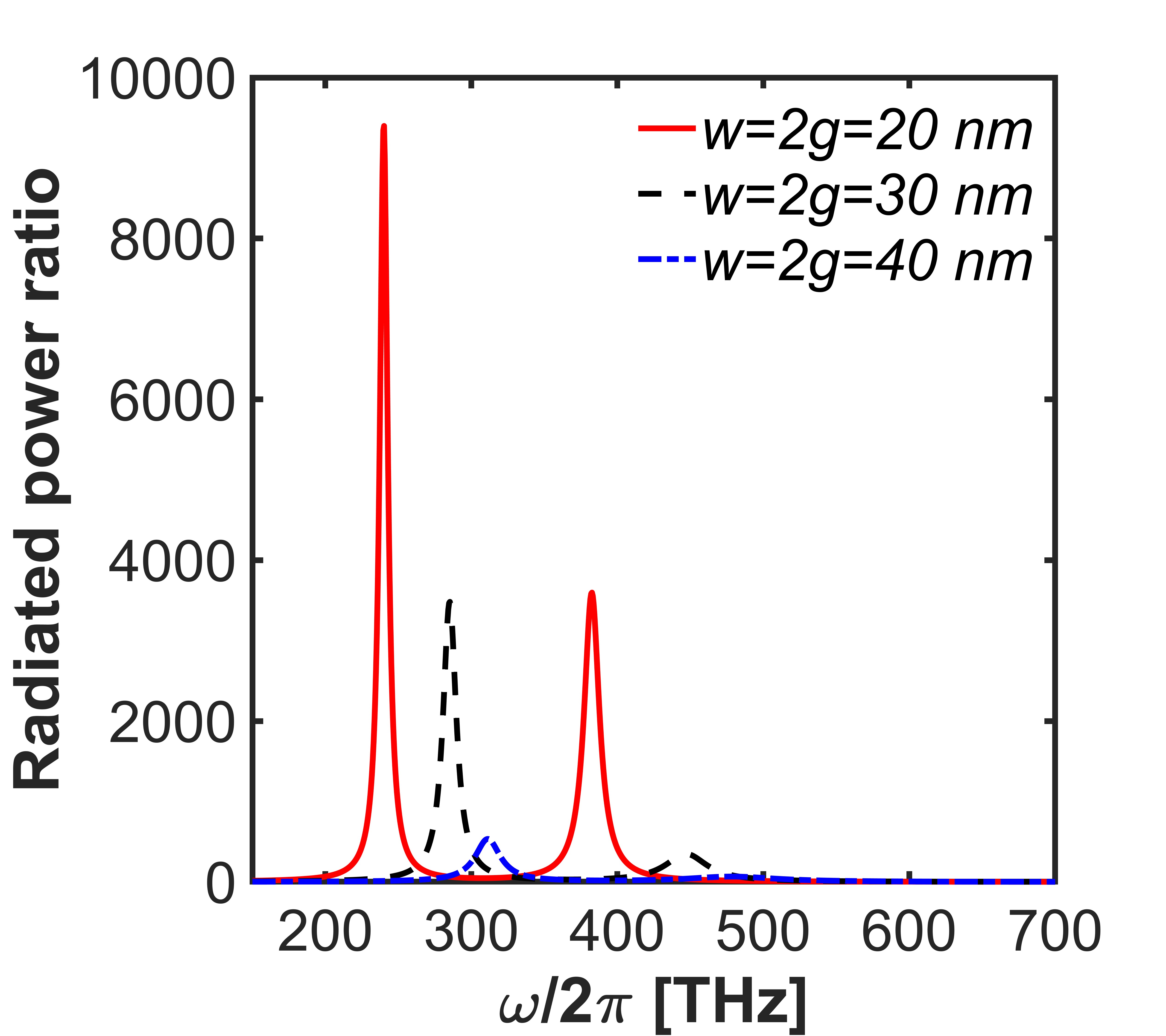} 
   \end{minipage}\hfill
   \begin{minipage}{0.25\textwidth}
       \centering
       \caption*{(c)}
       \includegraphics[width=\linewidth]{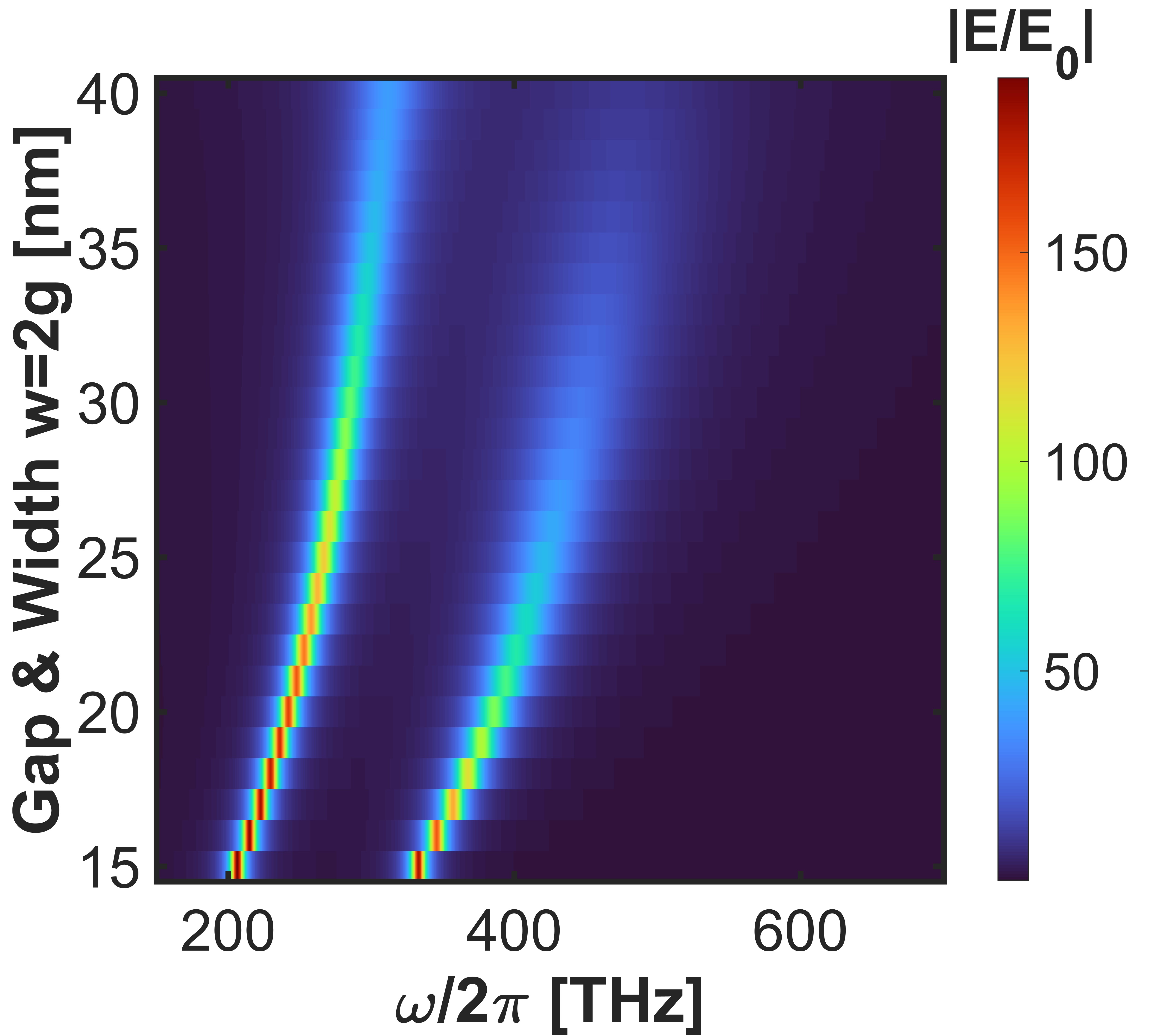} 
   \end{minipage}\hfill
   \begin{minipage}{0.25\textwidth}
       \centering
       \caption*{(d)}
       \includegraphics[width=\linewidth]{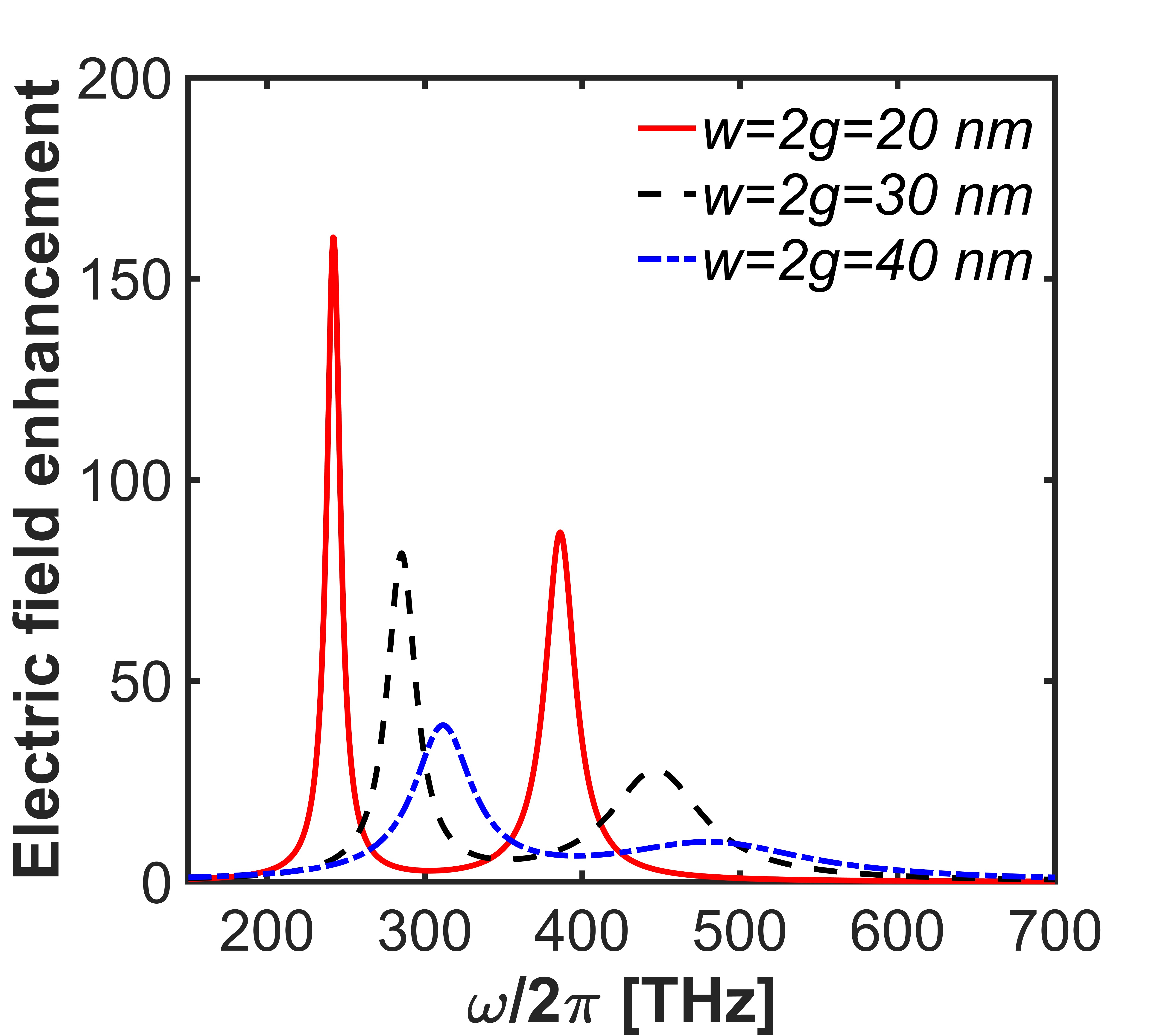} 
   \end{minipage}
   \caption{(a) Radiated power ratio, (b) radiated power ratio map by varying gap and width with mirror film. (c) Electric field enhancement, and (d) electric field enhancement map by varying gap and width in mirror film.}
   \label{fig:6}
\end{figure*}

The other parameters that influence the nanostructure's optical response are the width
and gap between the nanobars. In the considered plus-shaped silver nanobar configuration, the width is set as twice the gap, and for the simulation we varied $w=2g$ from $10\ \mathrm{nm}$ to $40\ \mathrm{nm}$. As shown in Fig.~\ref{fig:6}(a,b) and (c,d), increasing the values of $w$ and $g$ results in a reduction of the amplitude of the radiated power and electric field enhancement, accompanied by a blue-shift in the resonance spectrum. Additionally, as shown in Fig.~\ref{fig:6}(d), increasing the width and gap also broadens the resonance, indicating higher losses.
This will be important in our later considerations in \textit{section: Quantum simulations of two-photon absorption}.
Before we are ready to comment on this issue in a greater depth, we estimate the signal enhancement for specific molecules in a fully classical approach and compare it with the semiclassical one taking into account the saturation effects.

\section{TPA signal enhancement in molecules: classical approach}
\begin{figure*}[ht!]
   \centering
   \begin{minipage}{0.33\textwidth}
       \centering
       \caption*{(a)}
       \includegraphics[width=\linewidth]{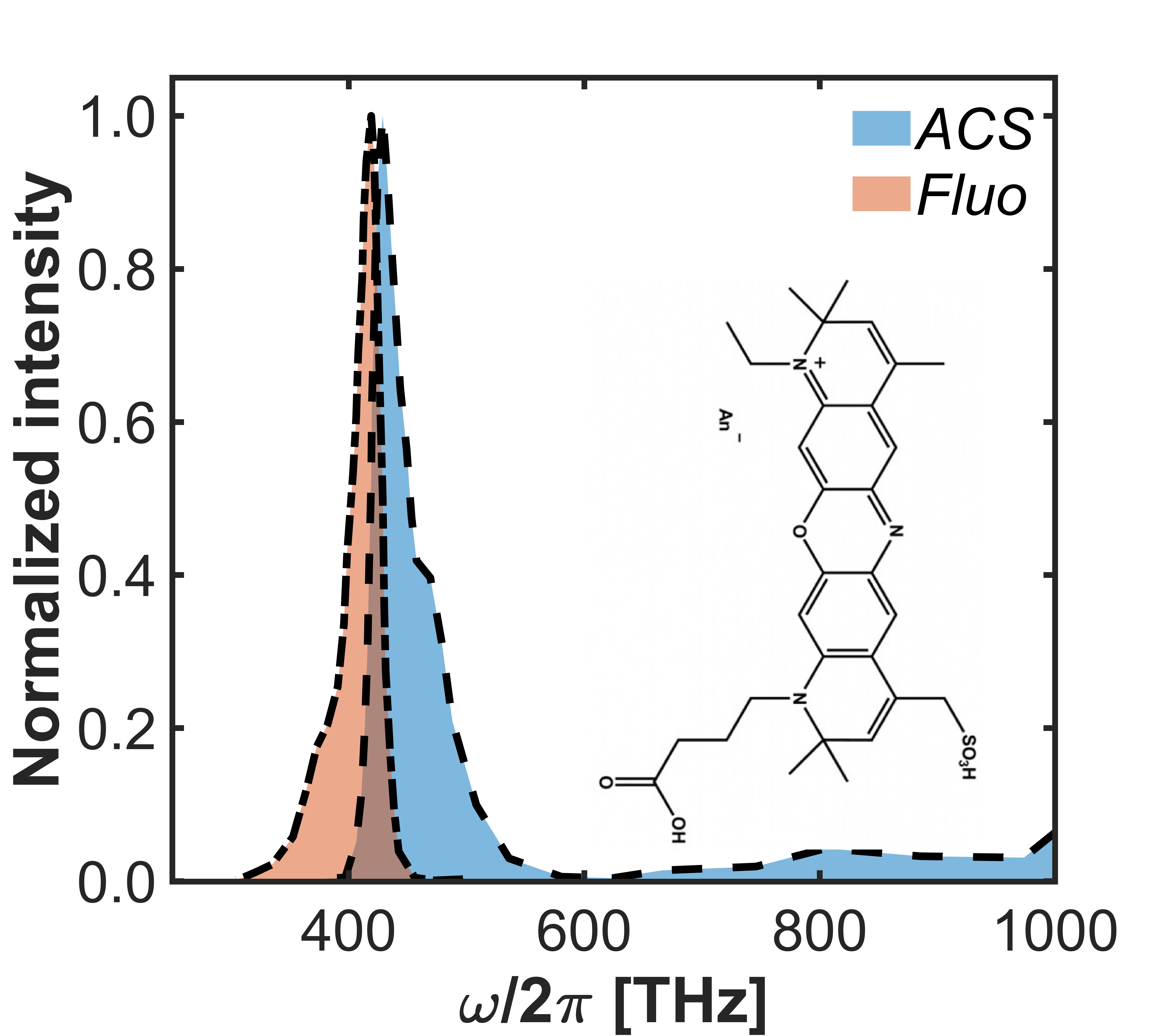}
   \end{minipage}\hfill
   \begin{minipage}{0.33\textwidth}
       \centering
       \caption*{(b)}
       \includegraphics[width=\linewidth]{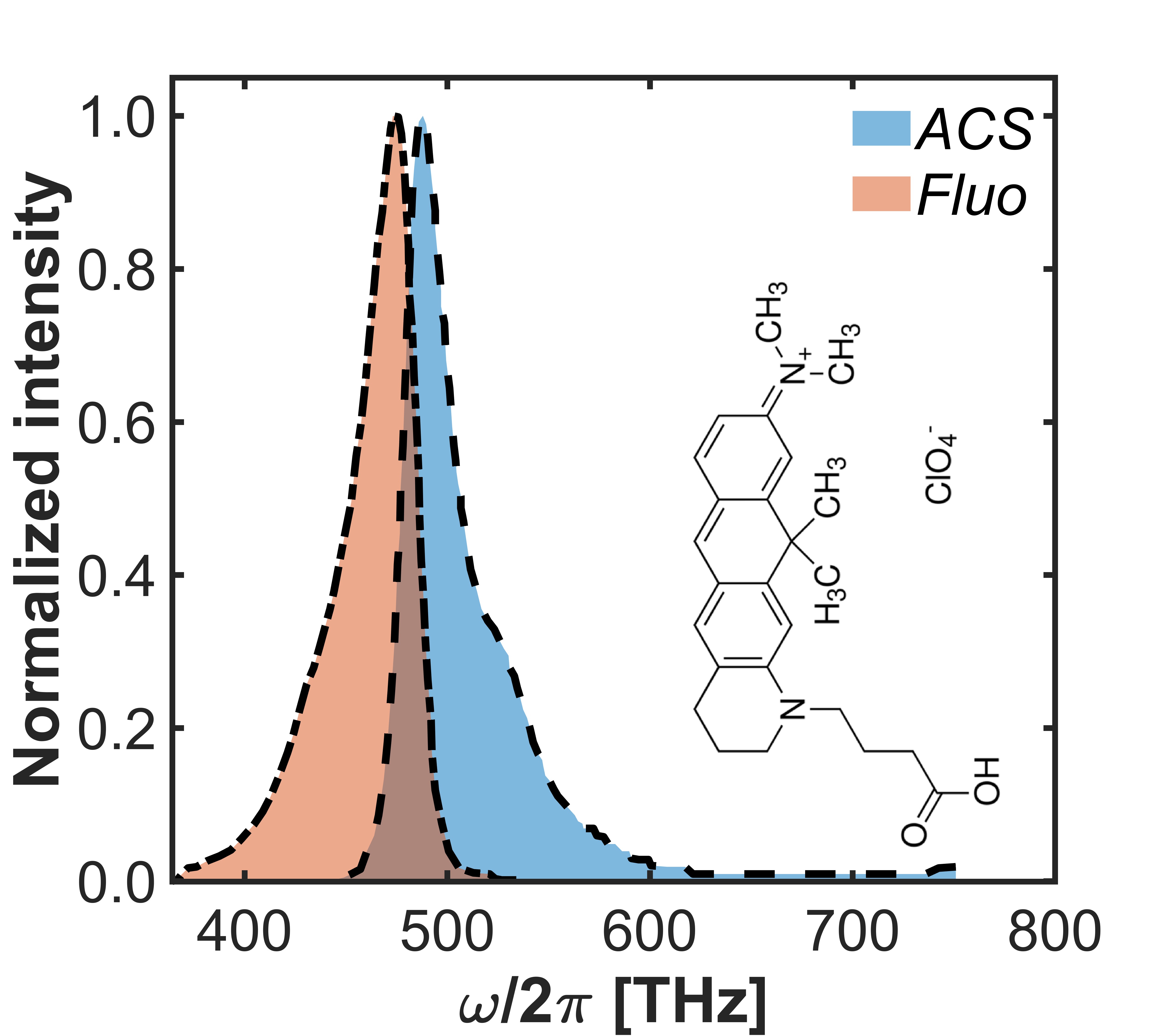}
   \end{minipage}\hfill
   \begin{minipage}{0.33\textwidth}
       \centering
       \caption*{(c)}
       \includegraphics[width=\linewidth]{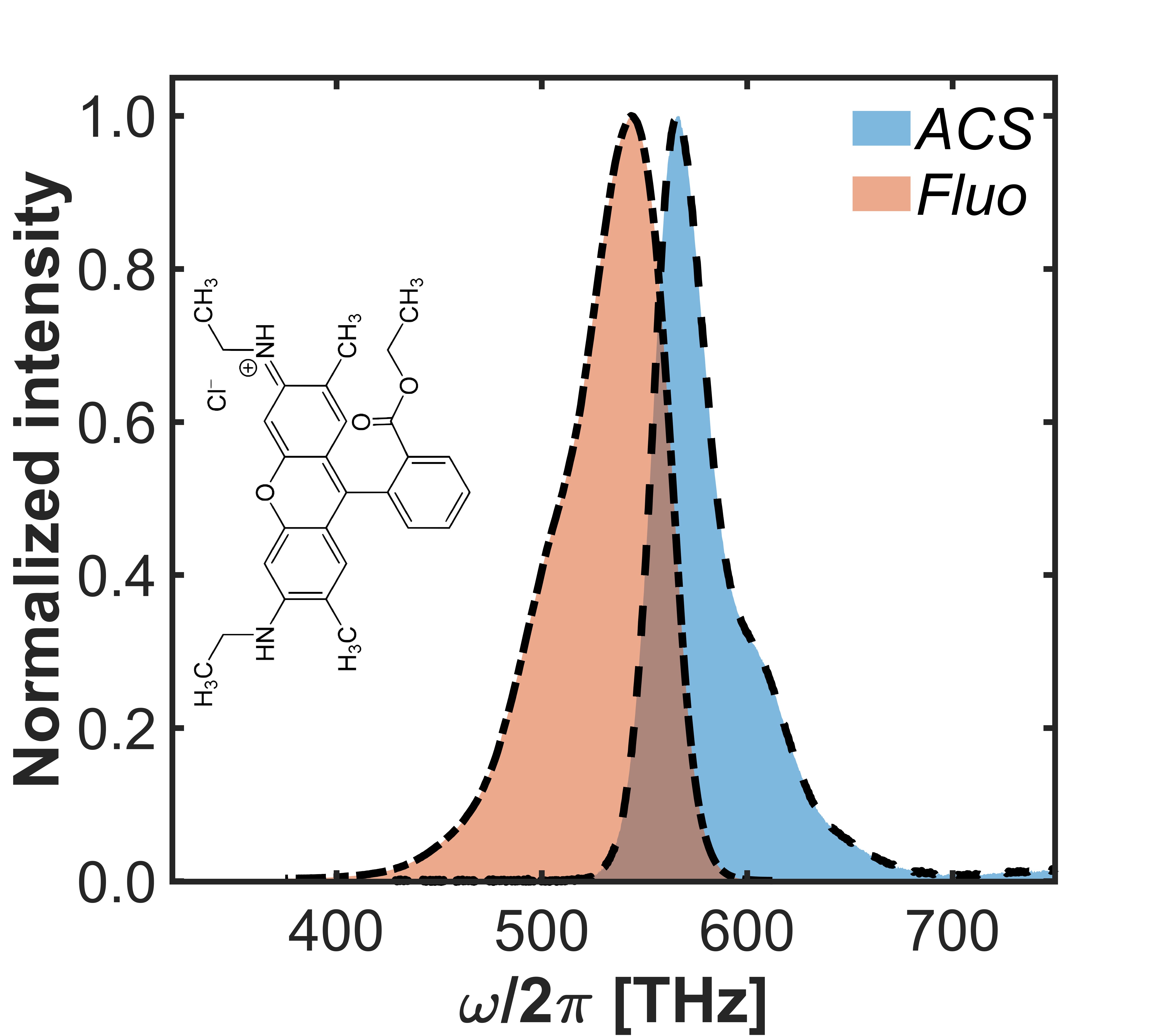}
   \end{minipage}
   \begin{minipage}{0.32\textwidth}
       \centering
       \caption*{(d)}
       \includegraphics[width=\linewidth]{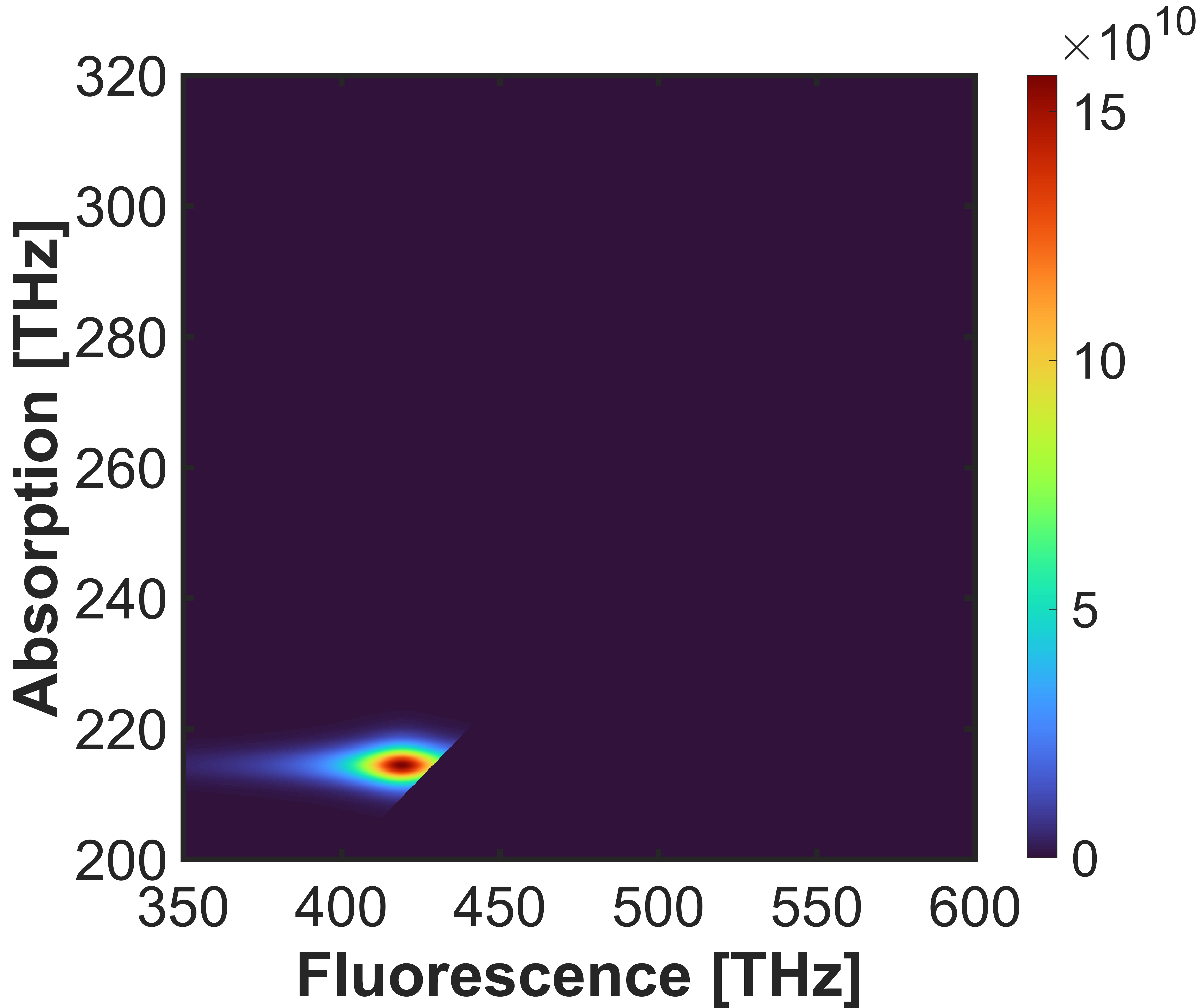}
   \end{minipage}\hfill
   \begin{minipage}{0.32\textwidth}
       \centering
       \caption*{(e)}
       \includegraphics[width=\linewidth]{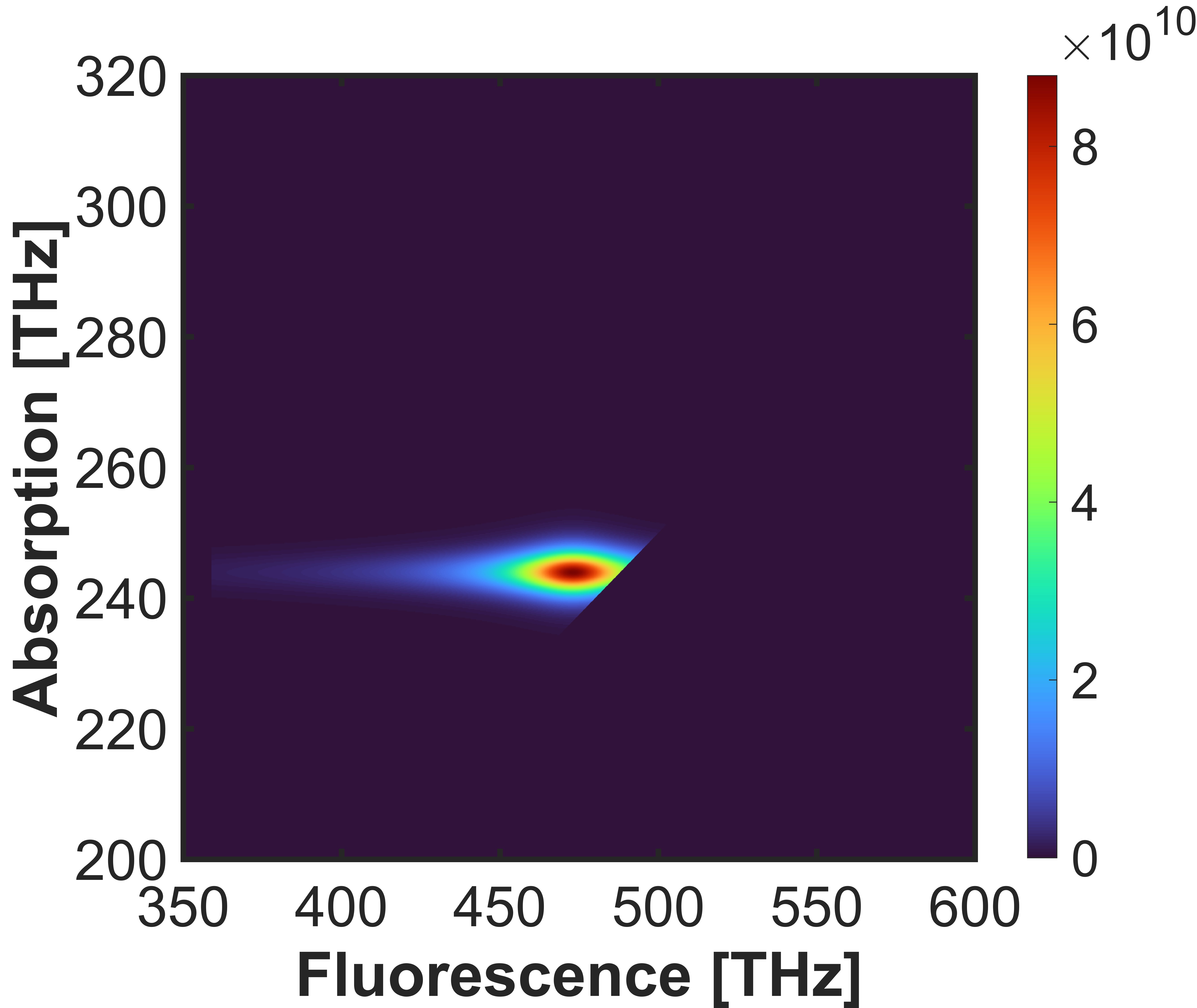}
   \end{minipage}\hfill
   \begin{minipage}{0.32\textwidth}
       \centering
       \caption*{(f)}
       \includegraphics[width=\linewidth]{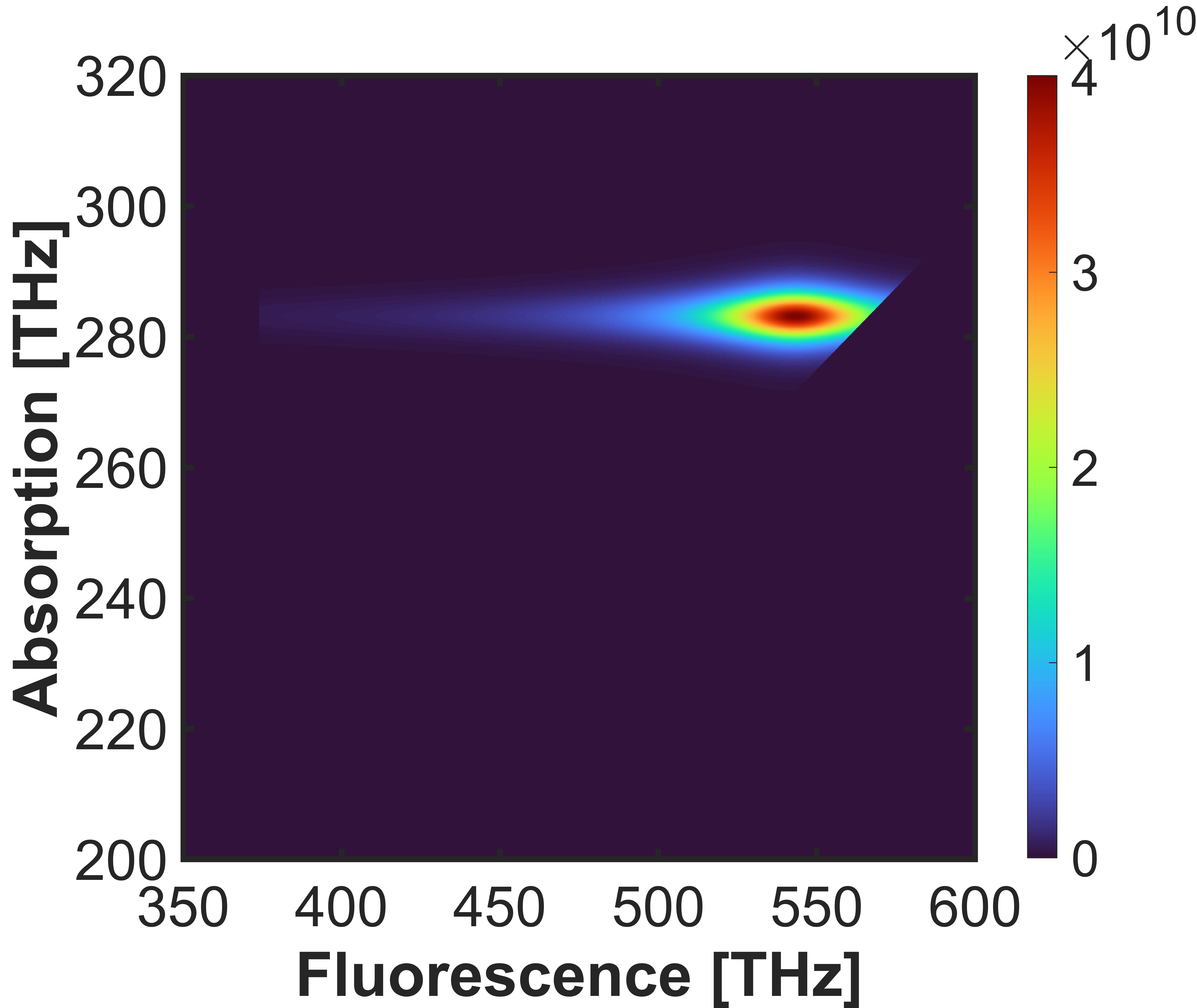}
   \end{minipage}
   \caption{Normalized intensity of fluorescence and absorption of (a) ATTO 700, (b) ATTO 610, and (c) Rho 6G. Signal enhancement map based on absorption and fluorescenece of (d) ATTO 700, (b) ATTO 610, and (c) Rho 6G.}
   \label{fig:7}
\end{figure*}

We now consider three different dye molecules: ATTO 700 \cite{AATBioSpectrum[Atto700]}, ATTO 610 \cite{AATBioSpectrum[Atto610]}, and Rho 6G \cite{AATBioSpectrum[AttoRho6G]}. Their single-photon absorption $C^{(1)}_\mathrm{abs}$ and fluorescence $F_\mathrm{flu}$ cross-sections are shown in Fig.~\ref{fig:7}(a-c). Asumming independent absorption and emission processes, the signal intensity component arising from subsequent absorption of a pair of photons at the illumination frequency $\omega_l$ and single-photon fluorescence at the frequency $\omega$ [Fig.~\ref{fig:1}(a)] is proportional to
\begin{equation}\label{eq:frequency-resolved-signal}
 I(\mathrm{r}_m,\omega_l,\omega) \propto \lvert E(\mathrm{r}_m,\omega_l)\rvert^{4} C_{abs}^{(2)}(\omega_l) P(\mathrm{r}_m,\omega) F_{flu}^{(1)}(\omega) \theta(2\omega_l-\omega),
\end{equation}
and, in general, depends on the molecular position with respect to the nanostructure. Here, $\theta(\cdot)$ is the Heaviside function that accounts for the assumption that the fluorescent photon cannot have energy larger than the absorbed pair. The total signal upon illumination with a monochromatic beam is thus \cite{izadshenas2024multiphoton}
\begin{equation}
 I(\mathrm{r}_m,\omega_l) \propto \lvert E(\mathrm{r}_m,\omega_l)\rvert^{4} C_{abs}^{(2)}(\omega_l) \int_0^{\omega_l} d\omega P(\mathrm{r}_m,\omega) F_{flu}^{(1)}(\omega).
\end{equation}
In both cases, in free space $E$ and $P$ should be replaced by $E_0$ and $P_0$.
Below, we evaluate frequency-resolved signal enhancement defined in the domain $2\omega_l\ge\omega$ as a ratio of expressions in Eq.~(\ref{eq:frequency-resolved-signal}) near the nanostructure and in free space
\begin{equation}\label{eq:signal_enhancement}
    I_\mathrm{se}(\omega_l,\omega) = \lvert\frac{E(\mathbf{r}_m,\omega_l)}{E_{0}(\omega_l)}\rvert^{4} \frac{P(\mathbf{r}_m,\omega)}{P_{0}(\omega)}
\end{equation}
Note that this definition is based on assumption that the excitation and emission processes are independent, hence, the overall enhancement can be evaluated as a product of enhancements of the two contributing processes. This approach is common in literature when considering plasmonic enhancement\cite{bharadwaj2009optical}. The validity of formula (\ref{eq:signal_enhancement}) and the underlying assumption in the context of TPA signal enhancement is verified in \textit{section: Quantum simulations of two-photon absorption}.

To demonstrate the nanostructure's tunability by design, we adjust the plasmonic nanostructure to each molecule separately. For the ATTO 700 dye molecule, the bar lengths were set to $L_{1} = 65.2 \ \mathrm{nm}$ and $L_{2} = 171.0 \ \mathrm{nm}$. For the ATTO 610 dye molecule, $L_{1} = 54.0 \ \mathrm{nm}$ and $L_{2} = 144.0 \ \mathrm{nm}$. For the Rho 6G dye molecule, $L_{1} = 42.7 \ \mathrm{nm}$ and $L_{2} = 116.3 \ \mathrm{nm}$. The remaining parameters remain as given in the previous section.
We match the plasmonic resonances with the two-photon absorption and fluorescence maxima of each dye molecule. For the two-photn absorption, we assume $C_{abs}^{(2)}(\omega)$ = $C_{abs}^{(1)}(2\omega)$ following experimental data\cite{drobizhev2011two} , which means that the NIR resonance is set at half the single-photon-absorption peak, and the visible resonance at the fluorescence peak. For different frequencies of the pump laser $\omega_l$ and the detection signal frequency $\omega$, we calculate the signal enhancement for each molecule according to Eq.~(\ref{eq:signal_enhancement}). The results are illustrated in Fig.~\ref{fig:7}(d-f), where we identify a resonant behaviour both in the absorption and fluorescence frequencies. The fluorescence resonance is broader due to the broader character of resonances at higher frequencies on the one hand, and to the 4th power with which the field enhancement is weighted on the other. Due to the high nonlinearity order, the overall frequency-resolved enhancement factor is predicted at large peak values of 10 to 11 orders of magnitude, and is slightly larger for molecules with fluorescence in lower frequency range. This greater enhancement is due to the plasmon frequency of silver providing more effective electric field confinement at lower frequencies. This result suggests that significant signal enhancement can be achieved for a molecules with the two-photon transition frequencies in a broad range within the visible regime.

So high predictions for the enhancements raise the question of the role of saturation, which is not accounted for in the above calculations. Below, we address this point by turning to the quantum description of molecular TPA in \textit{section: Semiclassical description of TPA}.

\section{Quantum simulations of two-photon absorption}
\label{sec:quantum}

In this section, we apply the quantum-mechanical description to evaluate the TPA signal enhancement and verify the assumption of independent enhancement of the excitation and the emission processes being the basis of Eq.~(\ref{eq:signal_enhancement}).

We consider a molecular model system with a single intermediate level detuned from the resonance with the illuminating beam, as shown in Fig.\ref{fig:1}(a). In this case, the sum in Eq.~(\ref{eq:Rabi}) is replaced by a single term. The calculations are performed with the Quantum Toolbox in Python package QuTiP \cite{johansson2012qutip}. For this section, we assume that the fluorescent light is emitted at doubled absorption frequency and the system is illuminated resonantly, i.e. $\omega = \omega_{eg} = 2\omega_l$. We detune the spectral position of the virtual state $\omega_{ig}$ by $\delta_{ig}=\omega_{ig}-\omega_l = 0.1\omega_{eg}$. 

The quantity of interest is the stationary signal enhancement, defined as the ratio of the number of photons emitted by the molecule in presence of the nanostructure and in free space.
\begin{equation}\label{eq:signal_enhancement_qm}
 I_\mathrm{qm} = \frac{\rho_{ee}^\mathrm{NP}\gamma}{\rho_{ee}^0\gamma_0}.
\end{equation}
The subscript "qm" stands for \textit{quantum mechanical}, "NP" denotes the case with the nanostructure and "0" - the free space scenario. The number of emitted photons is calculated as the product of the excited-state population $\rho_{ee}$ and the fluorescence rate $\gamma$, which may be Purcell-enhanced near the nanostructure.
Note that the enhancement calculated in this way accounts for the photons emitted by the molecule, but not for what happens with them afterwards: A photon could be radiated in the far field, where it can be detected to contribute to the signal, or absorbed by the metal forming the nanostructure. Thus, when estimating the signal one should rescale the above quantity by the nanostructure efficiency. In this section, however, we focus on mechanisms of photonic emission by the molecule and, hence, assume perfect efficiency for the moment.

\begin{figure*}[htbp]
    \begin{minipage}{0.33\columnwidth}
        \caption*{(a)}
        \includegraphics[width=\columnwidth]{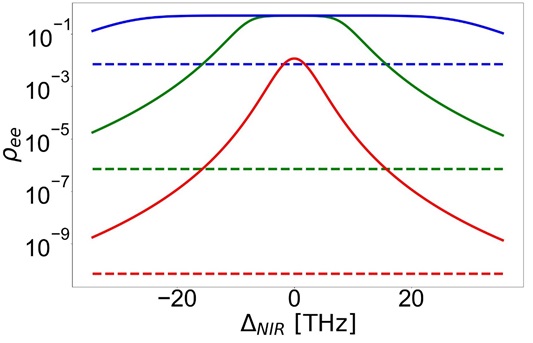}
    \end{minipage}\hfill
    \begin{minipage}{0.33\columnwidth}
        \caption*{(d)}
        \includegraphics[width=\columnwidth]{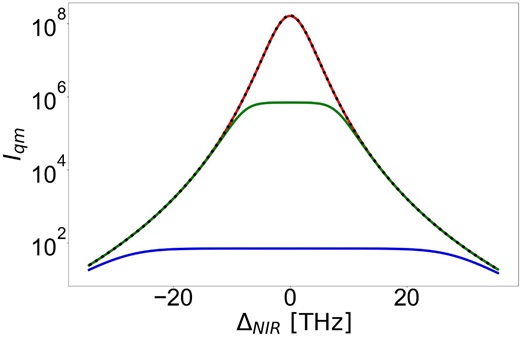}
    \end{minipage}
    \begin{minipage}{0.33\columnwidth}
        \caption*{(g)}
        \includegraphics[width=\columnwidth]{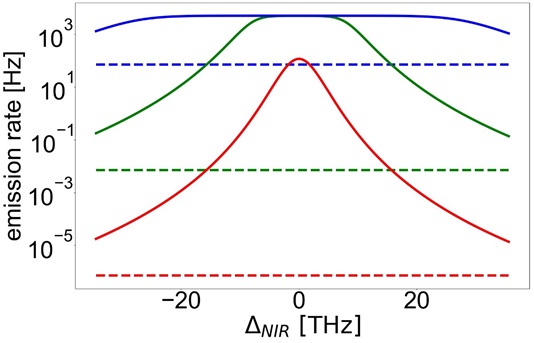}
    \end{minipage}\hfill
    \begin{minipage}{0.33\columnwidth}
        \caption*{(b)}
        \includegraphics[width=\columnwidth]{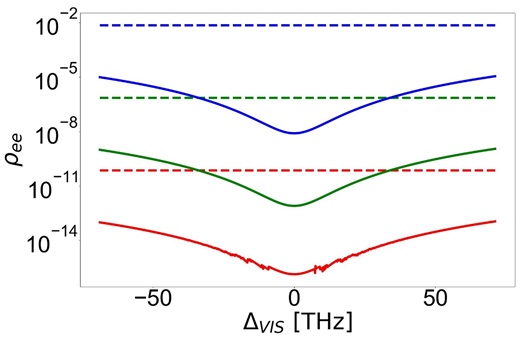}
    \end{minipage}\hfill
    \begin{minipage}{0.33\columnwidth}
        \caption*{(e)}
        \includegraphics[width=\columnwidth]{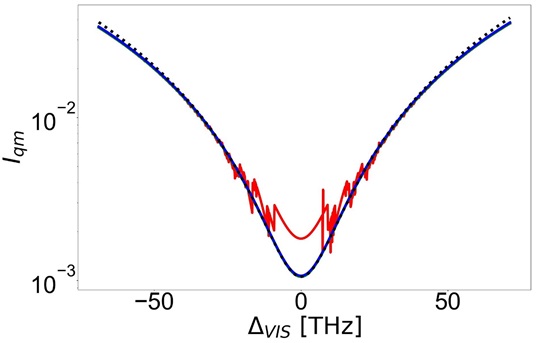}
    \end{minipage}\hfill
    \begin{minipage}{0.33\columnwidth}
        \caption*{(h)}
        \includegraphics[width=\columnwidth]{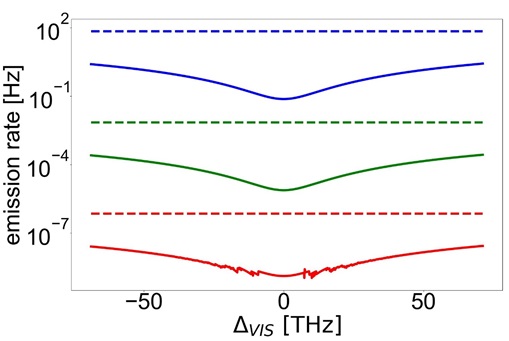}
    \end{minipage}\hfill
    \begin{minipage}{0.33\columnwidth}
        \caption*{(c)}
        \includegraphics[width=\columnwidth]{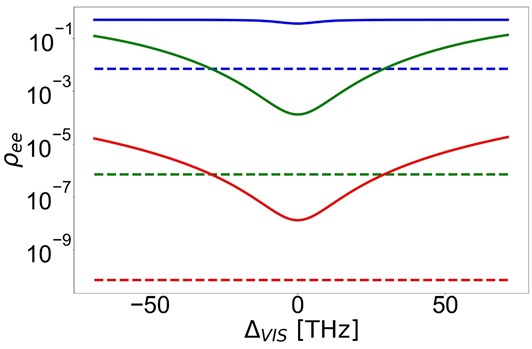}
    \end{minipage}\hfill
    \begin{minipage}{0.33\columnwidth}
        \caption*{(f)}
        \includegraphics[width=\columnwidth]{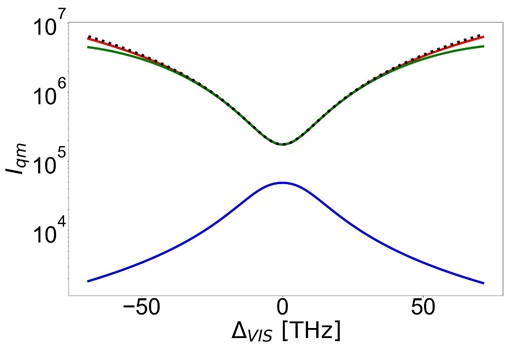}
    \end{minipage}\hfill
    \begin{minipage}{0.33\columnwidth}
        \caption*{(i)}
        \includegraphics[width=\columnwidth]{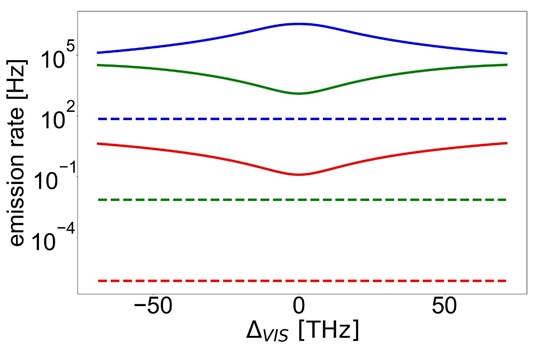}
    \end{minipage}
    \caption{Excited state population (a,b,c), TPA signal enhancement (d,e,f), and photon emission rate (g,h,i) in three scenarios. Panels (a,d,g) corresponds to the case of plasmonic NIR resonance tuned around the two-photon absorption resonance with the detuning $\Delta_\mathrm{NIR}$ from $\omega_{eg}/2$. Here, the visible resonance is assumed far-detund and the Purcell fluorescence enhancement factor is $1$.
    In (b,e,h), the NIR resonance is far detuned, with the local field enhancement factor being $1$, while the fluorescence is enhanced, as shown in function of the visible nanostructure resonance detuning $\Delta_\mathrm{VIS}$ from $\omega_{eg}$.
    In (e,f,i), both effects are simultaneously considered: the electric field is resonantly enhanced by the factor of 113.5, while the position of the visible resonance is modified according to the detuning $\Delta_\mathrm{VIS}$.
    In (a-c) and (g-i), solid lines correspond to the case of the molecule positioned near the nanostructure, while dashed lines - to the free-space case. In (d-f), solid lines are enhancement factors calculated quantum-mechanically, while black dotted lines denote the phenomenological enhancement factors: the fourth power of the field enhancement factor $E^4/E_0^4$ in (d), the inverse Purcell factor $1/F$ in (e), the inverse Purcell factor multiplied by the fourth power of the field enhancement factor $1/F \times E^4/E_0^4$ in (f). 
    In all panels, red lines are plotted for the illuminating beam amplitude leading to the single-photon interaction strength with the molecule of $g_0 = 1$ MHz, green lines - to $g_0 = 10$ MHz, and blue lines - to $g_0 = 100$ MHz.}
 \label{fig:qm}
\end{figure*}
The classical expression (\ref{eq:signal_enhancement}) is based on the assumption of the excitation and emission processes occuring at rates that can be independently enhanced. To verify this assumption in the TPA process, we compare $I_\mathrm{se}$ and $I_\mathrm{qm}$ in three cases: when two-photon absorption is enhanced by the nanostructure, but the fluorescent emission occurs at the free-space rate [Fig.~\ref{fig:qm}(a,d,g)], when the absorption occurs at the free-space rate, but fluorescence is Purcell-enhanced [Fig.~\ref{fig:qm}(b,e,h)], and when both contributions are plasmon-enhanced [Fig.~\ref{fig:qm}(c,f,i)].

We first analyze the case of enhanced excitation followed by nonehnanced emission ocurring at the free-space rate [Fig.~\ref{fig:qm}(a,d,g)]. This happens if the nanostructure's NIR resonance is tuned around the TPA absorption resonance of the molecule, while the VIS resonance is far detuned and the Purcell enhancement factor can be approximated as $1$. For the nanostructure field enhancement spectral profile, We assume a Lorentzian resonance centred at $\omega_\mathrm{NIR} = \Delta_\mathrm{NIR}+\omega_{eg}/2$ with the width of $2\pi\times 9.74$ THz, and present results as functions of the nanostructure NIR resonance detuning $\Delta_\mathrm{NIR}$.
We consider the process for three different amplitudes of the electric field of the illumination beam: $g_0 = 1$ MHz (red lines in all panels), $10$ MHz (green) and $100$ MHz (blue). In (a), stationary excited-state populations $\rho_{ee}$ due to TPA are shown with dashed lines in the free-space scenario. Naturally, in this case the population does not depend on the nanostructure detuning. The excited population grows with the illuminating-beam amplitude and, below the saturation level, scales with its fourth power.
The results obtained for the nanostructure vicinity are shown with solid lines. For the weakest of considered illumination strengths, the enhanced population is still below the saturation level and its spectral profile in (a) reflects the Lorentzian profile of the enhanced field. Thus, the enhancement factor $I_\mathrm{qm}$ evaluated according to Eq.~(\ref{eq:signal_enhancement_qm}) shows perfect agreement with the clasically evaluated enhancement $I_\mathrm{se}$ shown with the black dashed line in (d). The photon emission profile reflects the same shape in (g).
As the illumination strength is increased, a resonant nanostructure drives the molecular excited-state population at the saturation limit, as seen in a narrower range around the resonance for $g_0 = 10$ MHz (panel a, green), and a broader range for $g_0 = 100$ MHz (blue). As the saturation is reached, the enhancement factor drops (d) and spectral profile is modified with respect to the classical prediction. As a result, the photon emission rate reaches the saturation level in (g).
We conclude that the classical prediction for signal enhancement due to the plasmonically-enhanced field amplitude is valid below the saturation level.

We turn to the analysis of the impact of Purcell enhancement of the spontaneous emission. In all calculations, the free-space emission rate is assumed at the level of $\gamma_0 = 10$ kHz, corresponding to a relatively low single-particle transition dipole moment, as it might be realistic for molecules supporting two-photon absorption. Again, free-space population and photon emission rate levels are indicated with dashed lines in panels (b) and (h), respectively. For the Purcell-enhanced case, we assume the NIR resonance to be far-detuned so that the electric field enhancement factor is $1$, while the visible resonance position is sweeped according to the detuning $\Delta_\mathrm{VIS}$ of the resonance modeled with a Lorentzian lineshape with the width of $2\pi\times 23.1$ THz. We find that the Purcell-enhanced emission rate leads to a suppression of the excited-state population in (b). Below saturation, the impact of the Purcell enhancement has the same profile for all considered illumination amplitudes (red,green,blue). Thus, the Purcell effect \textit{decreases} the signal, as shown in panel (e). In the same panel, we also demonstrate that this result is reflected by the \textit{inverse} of the classical prediction: the black dotted line is the inverse $1/F$ of the Purcell enhancement factor evaluated as $F = P(\omega)/P_0(\omega)$. This result can be understood as follows: A large Purcell enhancement leads to emission of photons shortly after the excitation of the molecule. However, as the molecular excitation rate in TPA is small, further Purcell enhancement leads to an even faster photon emission each time a molecule gets excited, but the overall number of excitation events remains low. On the contrary, Purcell effect suppresses the excited-state probability and, as a result, the number of photons emitted per unit time.

Finally, we allow both the local electric field and the Purcell emission enhancements by both NIR and visible resonances in panels (c,f,i). We assume the NIR nanostructure resonance to be tuned exactly at half the molecular transition so that $\Delta_{NIR} = 0$. The visible resonance is tuned with the Lorentzian profile as before.
This time, due to the strong absorption enhancement, saturation is reached in the excited-state population in the case of strongest illumination beam (blue) in panel (c). Below the saturation limit, the red and green lines indicate excited-state population strongly increased with respect to the free-space level. As before, Purcell enhancement of spontaneous emission rate leads to a resonant dip in the stationary excited population. However, when the saturation level is reached, the Purcell enhancement may not be able to efficiently suppress the excited-state population as its impact is balanced by the enhanced field increasing the two-photon absorption level. As a result, we find the population steadily at the saturation limit as shown by the solid blue line in panel (c).
This qualitatively different behavior below and within the saturation limit is reflected in the signal enhancement factors in panel (f): the unsaturated results have the same profile as in panel (e), however, at a higher level due to the impact of the field enhancement. This result is overlapped with the black dotted line given by the product of the classical enhancement factor $(E/E_0)^4$ and the \textit{inverse} $P_0/P$ of the Purcell factor. However, in the saturation regime, the Purcell enhancement actually improves the signal. This time, the enhanced field drives the two-photon absorption so that the excitation rate keep up with the emission rate and, in consequence, the act of emitting a photon does not lead to a drop in the excited population level. Therefore, in the saturation regime a peak in the enhancement factor $I_\mathrm{qm}$ reappears: The blue solid line overlaps with the Purcell enhancement curve rescaled by the excited population ratio $\frac{P}{P_0}\frac{\rho_{ee}}{\rho_{ee}^0}$. The resulting photon emission rate is presented in panel (i) and again, shows qualitatively different profiles in the unsaturated (red, green lines) and saturated regimes (blue line).

In summary, analysing plasmonic TPA signal enhancement in the quantum-mechanical approach, we were not able to identify a case for which the molecular excitation and fluorescent emission could be described as independent - the assumption justifying Eq.~(\ref{eq:signal_enhancement}). On the contrary, we identify qualitatively different mechanisms for plasmonic signal enhancement to be efficient below and above the saturation limit. 
Below the saturation level, TPA signal can be enhanced through increasing the stationary probability for the molecule to be in the excited state. This can be achieved through electric field enhancement, fully in line with the classical predictions. Enhancing the fluorescent emission rate through the Purcell mechanism turns out inefficient, because it suppresses the probability of molecular excitation, and in consequence, the number of emitted photons. In other words, in the unsaturated case Purcell enhancement increases the rate at which a photon is emitted from an excited molecule, but does not increase the number of emitted photons. This is due to the excitation rate being too small in the unsaturated regime to keep up with the emission rate. 
In the saturation regime, on the other hand, field enhancement targeting the excitation stage is inefficient - the molecule is already saturated. In this case, excitation rate exceeds the emission rate and signal enhancement can be achieved by improving the latter. Hence, Purcell fluorescence enhancement does not lead to suppressing the excited-state population and becomes an efficient factor increasing the signal. These findings are summarized in Table 1.

\begin{table}[t!]
\begin{tabular}{l|c|c|c}
                      &  field  & fluorescence & phenomenological signal  \\
                      &  enhancement  & enhancement & enhancement formula \\
\hline
below saturation      &  efficient  &  counterproductive & $\frac{E^4}{E_0^4} \times \frac{1}{P/P_0}$\\
saturation regime     &  limited efficiency  & efficient  & $\frac{P}{P_0} \times \frac{\rho_{ee}}{\rho_{ee}^0}$
\end{tabular}
\caption{Summary of findings in Section \textit{Quantum simulations of two-photon absorption}. Saturation limit determines two distinct regimes where different enhancement scenarios should be targeted. }
\end{table}

The nanostructure design proposed in \textit{Section: Plasmonic nanostructure} could be used for verification of the above findings. The illuminating beam amplitude provides a control knob for reaching the saturation level, while its polarization can be a factor orienting the molecule in a selected direction for which one can target the two-photon absorption enhancement through local field confinement in the NIR mode, or fluorescence enhancement in the Purcell mechanism for the visible mode.

\section{Conclusions}
The quantum-mechanical analysis of plasmonic TPA signal enhancement indicates that molecular excitation and fluorescent emission are not independent processes. We identified distinct mechanisms for signal enhancement below and above the saturation limit. Below saturation, TPA signal enhancement is driven by increasing the stationary probability of the molecule being in the excited state, primarily through electric field enhancement. However, Purcell fluorescence enhancement is ineffective in this regime because it suppresses the excitation probability. Above saturation, where the molecule is already saturated, enhancing the emission rate becomes the key to increasing signal strength through Purcell enhancement. 

The proposed plus-shaped silver nanostructure design may offer a way to verify these findings. 
The plasmonic design has been fine-runed to match the optical characteristics of specific TPA-active dyes. The molecular absorption and fluorescence across near-infrared and visible spectra can be selectively addressed by illuminating beam polarization. 
Independent tuning of each plasmonic mode was achieved by adjusting nanobar lengths, which effectively controlled the resonance spectral characteristics. The integration of the metal mirror film further amplified the radiated power and electric field intensity through constructive interference. 

\section{Supplementary materials}

\subsection{Effective two-level description}\label{app:two-level}
Here, we derive the free-space form of the two-photon coupling strength in Eq.~(\ref{eq:Rabi}).

We consider a model molecule composed of ground and excited levels $|g\rangle,|e\rangle$ and a group of virtual levels $\{|i\rangle\}$ of energies $\omega_{g,e,\{i\}}$, respectively, and we assume $\omega_g<\omega_i<\omega_e$. The Bohr frequencies are $\omega_{mn}=\omega_m-\omega_n$ with $m,n\in\left\{g,e,\{i\}\right\}$. 
The free Hamiltonian is
\begin{equation}
    H_0 = \hbar\sum_{j=g,e,\{i\}} \omega_j|j\rangle\langle j|,
\end{equation}
where the set labeled with $\{i\}$ includes all virtual states.

The system is subject to a monochromatic illumination with a frequency $\omega_l \approx \frac{1}{2}\omega_{eg}$ near the two-photon resonance. The interaction Hamiltonian is
\begin{equation}
    H_\mathrm{int} = \sum_{i}\hbar\left[\left(\Omega_{ig}^{(1)}|g\rangle\langle i|e^{i\omega_l t}+\Omega_{ig}^{(1)\star}|i\rangle\langle g|e^{-i\omega_l t} \right) 
    + \left(\Omega_{ei}^{(1)}|i\rangle\langle e|e^{i\omega_l t}+\Omega_{ei}^{(1)\star}|e\rangle\langle i|e^{-i\omega_l t} \right)\right],
\end{equation}
where $\hbar\Omega_{pq}^{(1)} = \mathbf{E}_0(\omega_l,\mathbf{r}_m)\cdot \mathbf{d}_{pq}$ is the coupling strength of the free-space field with the amplitude $\mathbf{E}_0(\omega_l)$ evaluated at the molecular position with the molecular transition described with the dipole moment $\mathbf{d}_{pq}$. We have applied the electric-dipole and the rotating-wave approximations.

The intermediate levels are not, in general, spectrally positioned in the middle of the energy gap between the ground and excited states so that the single-photon resonance condition is generally not met. Here, we assume that the single-photon detuning between the field and the single-photon transitions $|g\rangle \leftrightarrow |i\rangle$, $|i\rangle \leftrightarrow |e\rangle$, is much larger than the two-photon detuning, but much smaller than the transition frequency scales:
$$|2\omega_l-\omega_{eg}| \ll |\omega_l-\omega_{ig}|,|\omega_l-\omega_{ei}| \ll \omega_l.$$
In this case, the population of the middle state is typically small, and the state can be adiabatically eliminated from the dynamics. Below, we perform the elimination to find the relation between the single-photon and two-photon transition rates. 

In the Schroedinger picture, the molecular state is a superposition $|\psi(t)\rangle = c_g(t)|g\rangle+ c_e(t)|e\rangle+\sum_ic_i(t)|i\rangle$ with $c_g(0)=1$ and $|c_g(t)|^2+|c_e(t)|^2+\sum_i|c_i(t)|^2=1$. 
The Schr\"{o}dinger equation yields the following set of equations for the amplitudes:
\begin{eqnarray}
    i\dot{d_g} &=& \sum_i \Omega_{ig}^{(1)} d_i\nonumber\\
    i\dot{d_e} &=& (\omega_{eg}-2\omega_l)d_e+\sum_i \Omega_{ei}^{(1)\star}d_i\label{eq:schroedinger}\\
    i\dot{d_i} &=& (\omega_{ig}-\omega_l)d_i+ \Omega_{ig}^{(1)\star}d_g+\Omega_{ei}^{(1)} d_e,\nonumber
\end{eqnarray}
where we have set $\omega_g=0$ and substituted $c_e=d_ee^{-2i\omega_l t}$, $c_i=d_ie^{-i\omega_l t}$, $c_g\equiv d_g$ to approximately separate the free evolution.

Near the two-photon resonance, the virtual-level amplitudes are relatively small $|c_i(t)|\ll |c_g(t)|,|c_e(t)|$ and these states can be adiabatically eliminated from the dynamics
\begin{equation}
    d_i=\frac{\Omega_{ig}^{(1)\star}d_g+\Omega_{ei}^{(1)} d_e}{\omega_l-\omega_{ig}},
\end{equation}
which we plug back into the first pair of equations (\ref{eq:schroedinger}). 
We find an effective two-level system description
\begin{eqnarray}
i\dot{d_{g}} &=& \sum_{i}\frac{\left|\Omega^{(1)}_{ig}\right|^{2}}{\omega_l-\omega_{ig}}d_{g}+\sum_{i}\frac{\Omega^{(1)}_{ig}\Omega^{(1)}_{ei}}{\omega_l-\omega_{ig}}d_{e}, \\
i\dot{d_{e}} &=& \left(\omega_{eg}-2\omega_l+\sum_{i}\frac{\left|\Omega^{(1)}_{ei}\right|^{2}}{\omega_l-\omega_{ig}}\right)d_{e}+\sum_{i}\frac{{\Omega^{(1)}}^\star_{ei}{\Omega^{(1)}}^\star_{ig}}{\omega_l-\omega_{ig}}d_{g}.
\end{eqnarray}
In the above equations, we identify frequency shifts $\Delta\omega_j=\sum_i\frac{|\Omega_{ij}|^2}{\omega_l-|\omega_{ij}|}$, $j\in\{e,g\}$ 
and the effective coupling constant between the ground and excited states 
\begin{equation}\label{eq:effective_coupling}
    \Omega^{(2)} = \sum_i\frac{\Omega^{(1)}_{ig}\Omega^{(1)}_{ei}}{\omega_l-\omega_{ig}}.
\end{equation} 
The effective Hamiltonian takes the form given in Eq.~(\ref{eq:hamiltonian}).
This result justifies the two-level model assumed in \textit{section: Semiclassical description of TPA}.

To account for the presence of the plasmonic nanostructure, each of the contributions of the effective coupling strength (\ref{eq:effective_coupling}) arising due to the presence of the intermediate states is rescaled by factors related to the plasmonic enhancement of the field component parallel to the corresponding electric dipole moment elements $\mathbf{d}_{ig}$ and $\mathbf{d}_{ei}$ of the transitions involving the virtual states:
$$\Omega^{(2)}_\mathrm{NP}=\sum_i\underbrace{\frac{\mathbf{E}(\omega_l,\mathbf{r}_m)\cdot \mathbf{d}_{ei}}{\mathbf{E}_{0}(\omega_l)\cdot \mathbf{d}_{ei}}
\frac{\mathbf{E}(\omega_l,\mathbf{r}_m)\cdot \mathbf{d}_{ig}}{\mathbf{E}_{0}(\omega_l)\cdot \mathbf{d}_{ig}}}_\mathrm{enhancement\ factors}
\underbrace{\frac{\Omega^{(1)}_{ig}\Omega^{(1)}_{ei}}{\omega_l-\omega_{ig}}}_\mathrm{free-space\ two-photon\ coupling}.$$
Assuming all transition dipoles to be co-oriented, the two-photon-transition coupling strength with the external field becomes rescaled by the field enhancement factors, as given in Eq.~(\ref{eq:Rabi}). 

\subsection{Plasmonic nanostructure without mirror film}
\begin{figure*}[ht!]
   \centering
   \begin{minipage}{0.25\textwidth}
       \centering
       \caption*{(a)}
       \includegraphics[width=\linewidth]{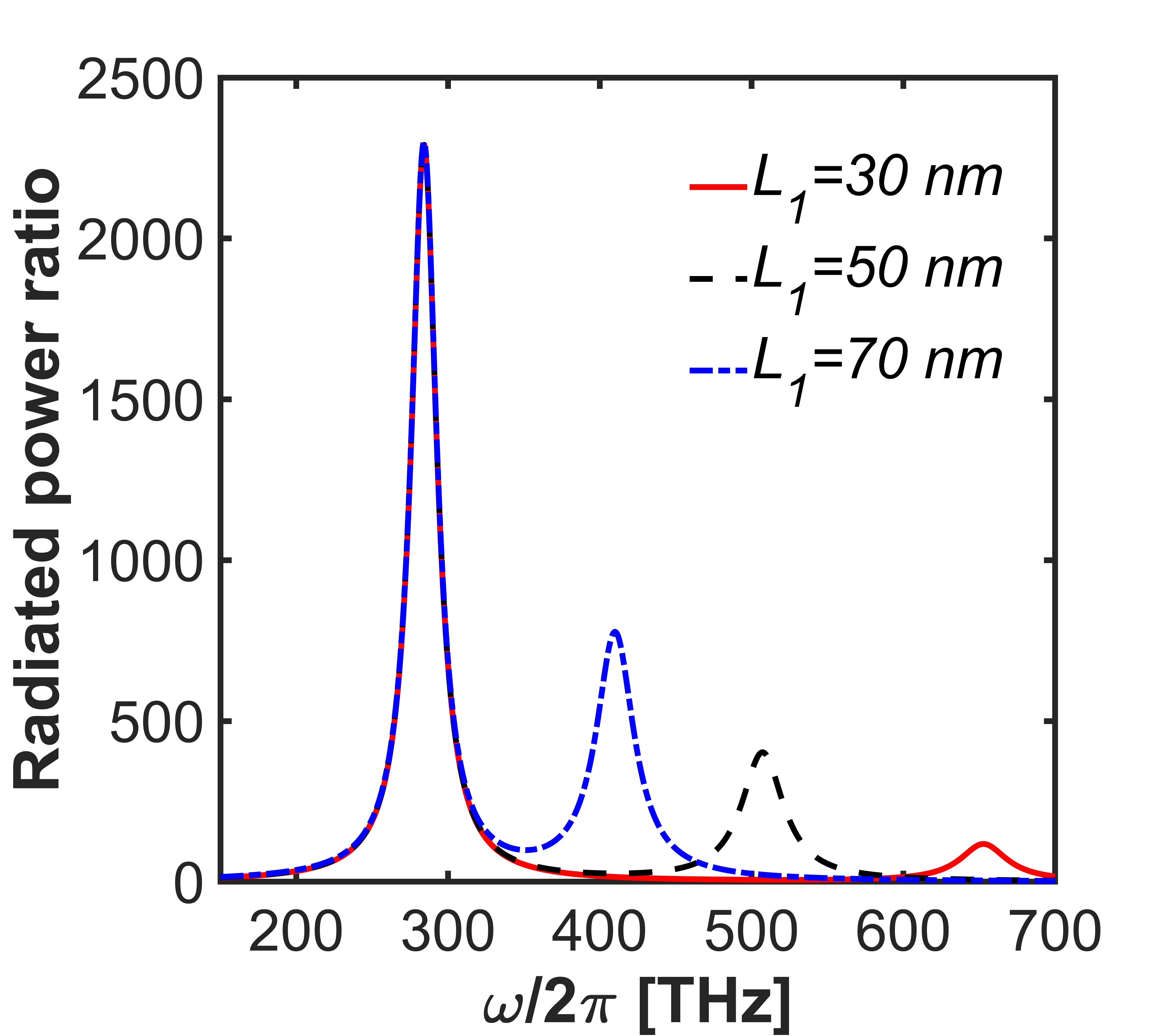}
   \end{minipage}\hfill
   \begin{minipage}{0.25\textwidth}
       \centering
       \caption*{(b)}
       \includegraphics[width=\linewidth]{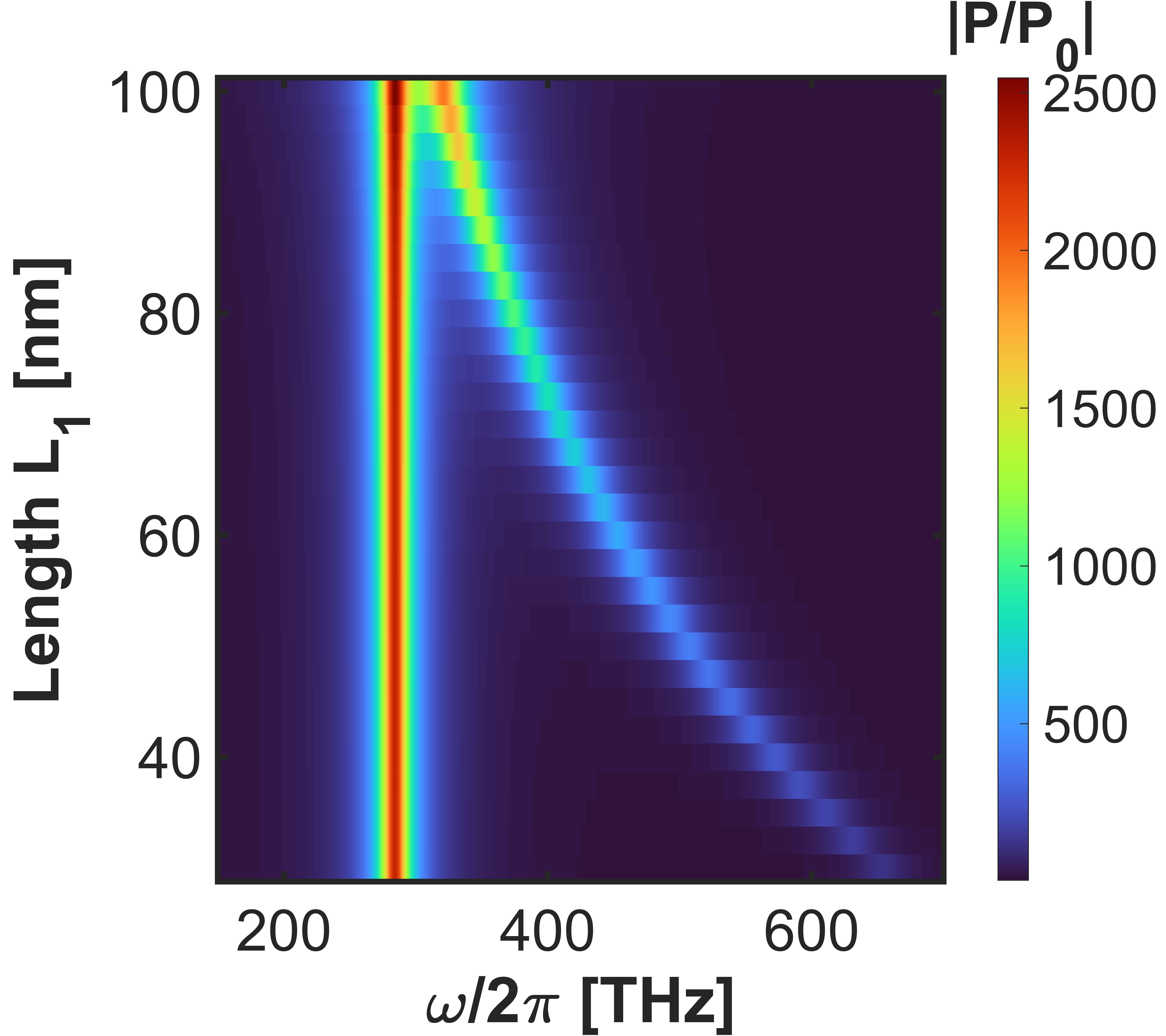}
   \end{minipage}\hfill
   \begin{minipage}{0.25\textwidth}
       \centering
       \caption*{(c)}
       \includegraphics[width=\linewidth]{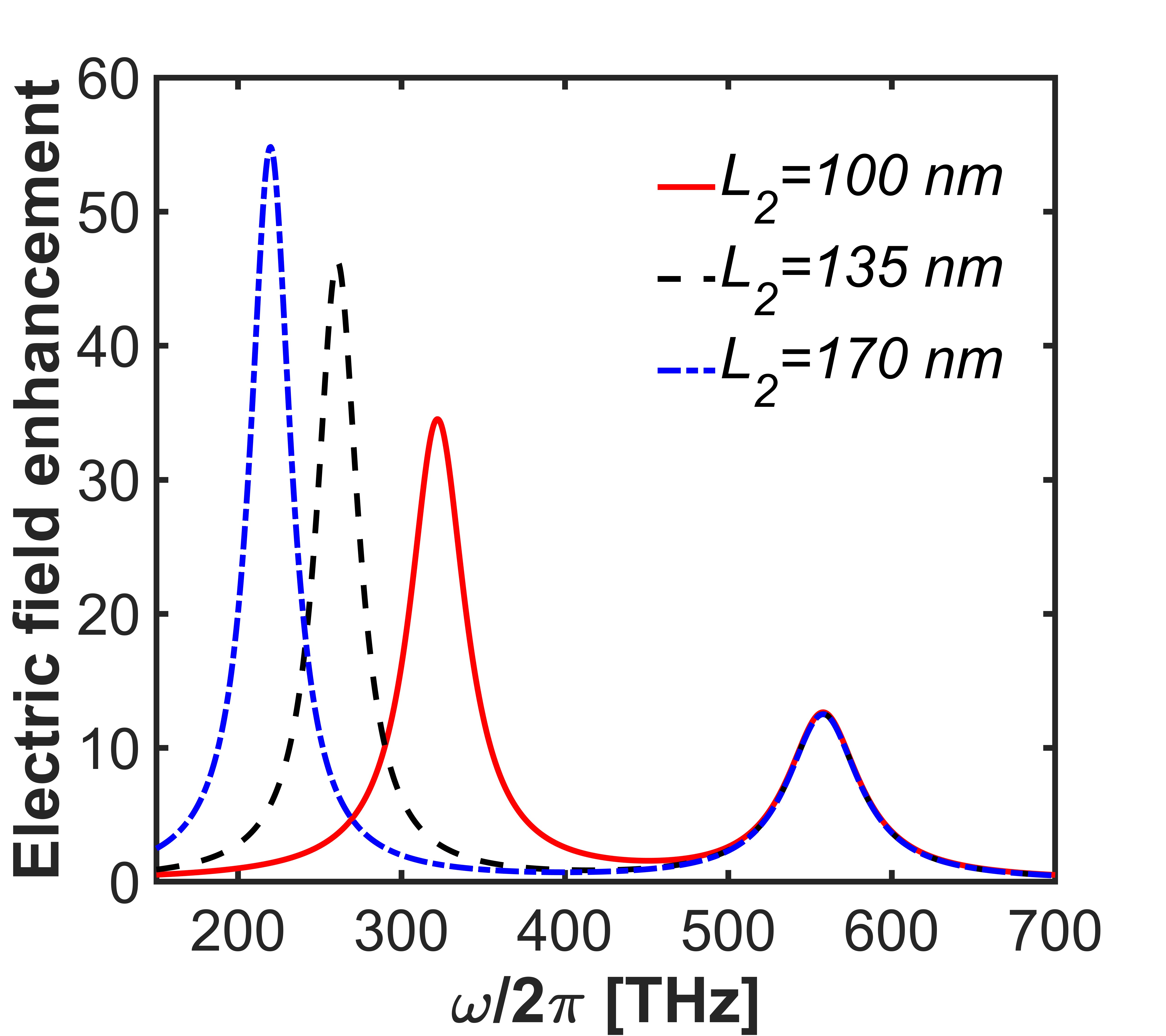}
   \end{minipage}\hfill
   \begin{minipage}{0.25\textwidth}
       \centering
       \caption*{(d)}
       \includegraphics[width=\linewidth]{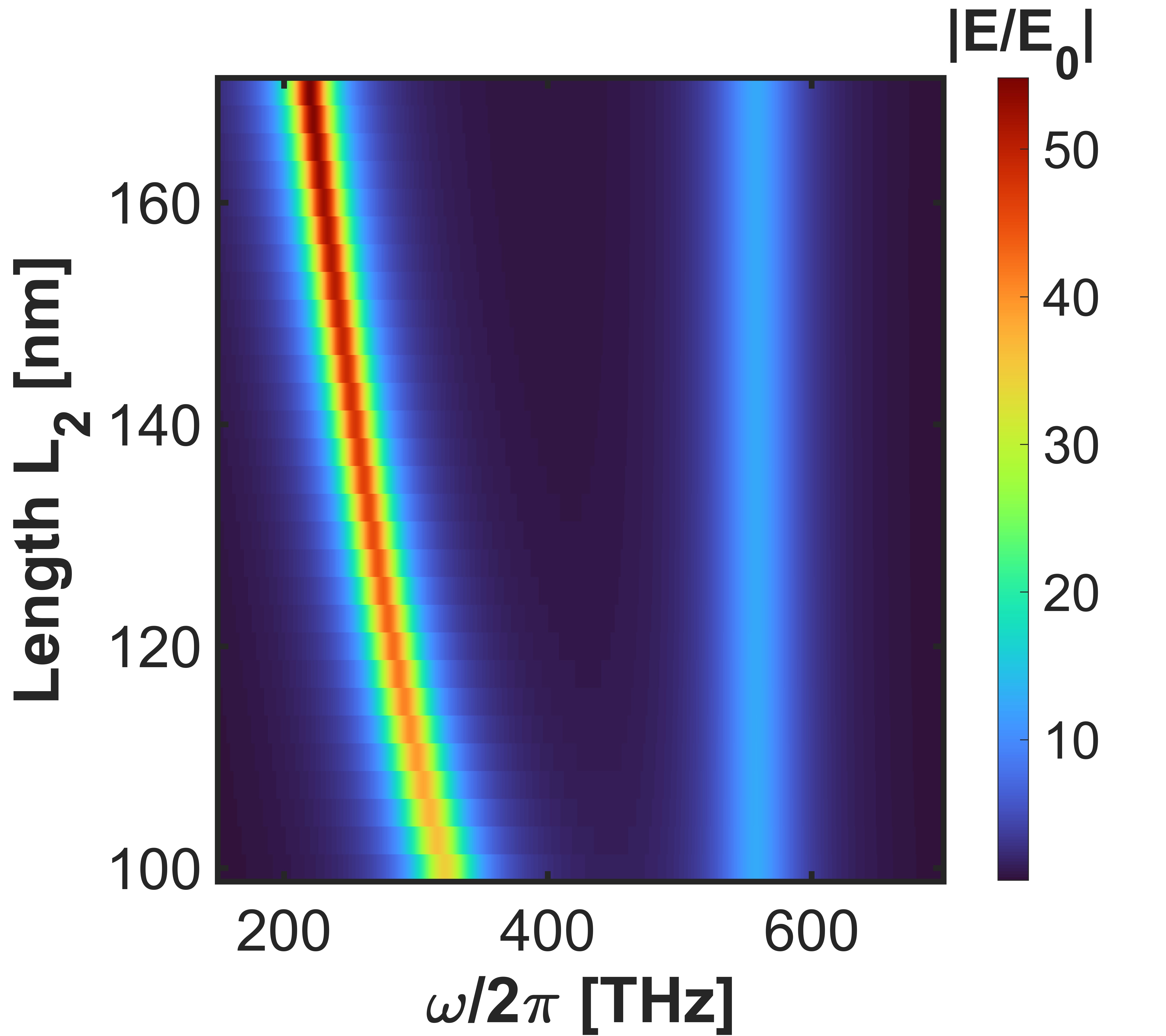}
   \end{minipage}
   \caption{(a) Radiated power for selected lengths, and (b) radiated power spectrum by varying $L_{1}$  (c) Electric field enhancement, and (d) electric field enhancement spectrum by varying $L_{2}$.}
   \label{fig:2}
\end{figure*}

\section{Methods}

\subsection{Classical electromagnetic simulations}
We numerically model the optical response of the nanostructure using the finite integration method in COMSOL Multiphysics. We employed the user-control mesh method to effectively adjust the mesh size according to the element present in the simulated structure. The silver nanostructures model with the Drude model fits to dielectric permittivity based on experimental data of Johnson and Christy \cite{johnson1972optical}. The refractive index of silica is $n=1.45$  \cite{malitson1965interspecimen}.

The results have been obtained for separately modeled infrared to visible problems, with identical, COMSOL-built-in scattering boundary conditions at a $1\ \mu$m-diameter sphere that prevents reflection back from infinite space, with an additional perfectly matched layer.

We use Poynting's theorem to calculate the power radiated from and absorbed by the nanostructure  \cite{jackson,izadshenas2023hybrid,izadshenas2023metasurface}:
\begin{eqnarray}
    P_\mathrm{rad}(\omega) &=&  \oint\ \langle\vec{E}_\mathrm{sca}(\vec{r},\omega) \times \vec{H}_\mathrm{sca}(\vec{r},\omega)\rangle \ d\vec{A}, \\
    P_\mathrm{abs}(\omega) &=&  \int\ \langle\vec{J}_\mathrm{ind}(\vec{r},\omega) \cdot \vec{E}_\mathrm{ind}(\vec{r},\omega)\rangle \ dV, 
\label{eq.S1}
\end{eqnarray}
where similarly to $\vec{E}_\mathrm{sca}(\vec{r},\omega)$, the symbol $\vec{H}_\mathrm{sca}(\vec{r},\omega)$ denotes the scattered part of the magnetic field, and $\vec{J}_\mathrm{ind}(\vec{r},\omega)$ and $\vec{E}_\mathrm{ind}(\vec{r},\omega)$ represent the currents and electric field in the nanostructure volume. The integrals are evaluated, respectively, at the spherical surface of the simulation volume and inside the volume of the nanostructure elements.

\subsection{Two-photon absorption calculations}
For our quantum calculations, we used the QuTiP package in Python with the following details: The longer nanobars have a length of $L_{2}=171\ \mathrm{nm}$, with a frequency resonance at $214.42 \ \mathrm{THz}$ and an electric field enhancement of 113.49. The smaller nanobars are $L_{1}=65.3\ \mathrm{nm}$ in length, with a frequency resonance at $418.98\ \mathrm{THz}$ and a radiated power ratio of 948. The Rabi frequency of the molecule without the nanostructure is $2\pi\times 10^{10} \ \mathrm{[Hz]}$.

\begin{acknowledgement}

The authors acknowledge the funding by the National Centre for Research and Development, Poland, within the QUANTERA II
Programme under Project QUANTERAII/1/21/E2TPA/2023.

\end{acknowledgement}

\begin{suppinfo}

Derivation of effective two-level description.
Simulations of optical response of the plasmonic nanostructure without the mirror film. 

\end{suppinfo}

\bibliography{achemso-demo}

\providecommand{\latin}[1]{#1}
\makeatletter
\providecommand{\doi}
  {\begingroup\let\do\@makeother\dospecials
  \catcode`\{=1 \catcode`\}=2 \doi@aux}
\providecommand{\doi@aux}[1]{\endgroup\texttt{#1}}
\makeatother
\providecommand*\mcitethebibliography{\thebibliography}
\csname @ifundefined\endcsname{endmcitethebibliography}  {\let\endmcitethebibliography\endthebibliography}{}
\begin{mcitethebibliography}{35}
\providecommand*\natexlab[1]{#1}
\providecommand*\mciteSetBstSublistMode[1]{}
\providecommand*\mciteSetBstMaxWidthForm[2]{}
\providecommand*\mciteBstWouldAddEndPuncttrue
  {\def\EndOfBibitem{\unskip.}}
\providecommand*\mciteBstWouldAddEndPunctfalse
  {\let\EndOfBibitem\relax}
\providecommand*\mciteSetBstMidEndSepPunct[3]{}
\providecommand*\mciteSetBstSublistLabelBeginEnd[3]{}
\providecommand*\EndOfBibitem{}
\mciteSetBstSublistMode{f}
\mciteSetBstMaxWidthForm{subitem}{(\alph{mcitesubitemcount})}
\mciteSetBstSublistLabelBeginEnd
  {\mcitemaxwidthsubitemform\space}
  {\relax}
  {\relax}

\bibitem[Garcia‐Lechuga \latin{et~al.}(2014)Garcia‐Lechuga, Fuentes, Grützmacher, Pérez, and Rosa]{Garcia‐Lechuga2014Calculation}
Garcia‐Lechuga,~M.; Fuentes,~L.~M.; Grützmacher,~K.; Pérez,~C.; Rosa,~M. I. D.~L. Calculation of the spatial resolution in two-photon absorption spectroscopy applied to plasma diagnosis. \emph{Journal of Applied Physics} \textbf{2014}, \emph{116}, 133103\relax
\mciteBstWouldAddEndPuncttrue
\mciteSetBstMidEndSepPunct{\mcitedefaultmidpunct}
{\mcitedefaultendpunct}{\mcitedefaultseppunct}\relax
\EndOfBibitem
\bibitem[Yi \latin{et~al.}(2014)Yi, Yang, Peng, Liu, Li, Zhong, Yang, and Tan]{Yi2014Two-photon}
Yi,~M.; Yang,~S.; Peng,~Z.; Liu,~C.; Li,~J.; Zhong,~W.; Yang,~R.; Tan,~W. Two-photon graphene oxide/aptamer nanosensing conjugate for in vitro or in vivo molecular probing. \emph{Analytical chemistry} \textbf{2014}, \emph{86 7}, 3548--54\relax
\mciteBstWouldAddEndPuncttrue
\mciteSetBstMidEndSepPunct{\mcitedefaultmidpunct}
{\mcitedefaultendpunct}{\mcitedefaultseppunct}\relax
\EndOfBibitem
\bibitem[Pascal \latin{et~al.}(2021)Pascal, David, Andraud, and Maury]{Pascal2021Near-infrared}
Pascal,~S.; David,~S.; Andraud,~C.; Maury,~O. Near-infrared dyes for two-photon absorption in the short-wavelength infrared: strategies towards optical power limiting. \emph{Chemical Society reviews} \textbf{2021}, \emph{50 11}, 6613--6658\relax
\mciteBstWouldAddEndPuncttrue
\mciteSetBstMidEndSepPunct{\mcitedefaultmidpunct}
{\mcitedefaultendpunct}{\mcitedefaultseppunct}\relax
\EndOfBibitem
\bibitem[Ojambati \latin{et~al.}(2020)Ojambati, Chikkaraddy, Deacon, Huang, Wright, and Baumberg]{Ojambati2020Efficient}
Ojambati,~O.; Chikkaraddy,~R.; Deacon,~W.~M.; Huang,~J.; Wright,~D.; Baumberg,~J. Efficient Generation of Two-Photon Excited Phosphorescence from Molecules in Plasmonic Nanocavities. \emph{Nano Letters} \textbf{2020}, \emph{20}, 4653 -- 4658\relax
\mciteBstWouldAddEndPuncttrue
\mciteSetBstMidEndSepPunct{\mcitedefaultmidpunct}
{\mcitedefaultendpunct}{\mcitedefaultseppunct}\relax
\EndOfBibitem
\bibitem[Yang \latin{et~al.}(2015)Yang, Wang, Boulesbaa, Kravchenko, Briggs, Puretzky, Geohegan, and Valentine]{yang2015nonlinear}
Yang,~Y.; Wang,~W.; Boulesbaa,~A.; Kravchenko,~I.~I.; Briggs,~D.~P.; Puretzky,~A.; Geohegan,~D.; Valentine,~J. Nonlinear Fano-resonant dielectric metasurfaces. \emph{Nano letters} \textbf{2015}, \emph{15}, 7388--7393\relax
\mciteBstWouldAddEndPuncttrue
\mciteSetBstMidEndSepPunct{\mcitedefaultmidpunct}
{\mcitedefaultendpunct}{\mcitedefaultseppunct}\relax
\EndOfBibitem
\bibitem[Lee \latin{et~al.}(2001)Lee, Lee, Kim, Choi, Cho, Jeon, and Cho]{Lee2001Two-photon}
Lee,~W.; Lee,~H.; Kim,~J.-A.; Choi,~J.~H.; Cho,~M.; Jeon,~S.; Cho,~B.~R. Two-photon absorption and nonlinear optical properties of octupolar molecules. \emph{Journal of the American Chemical Society} \textbf{2001}, \emph{123 43}, 10658--67\relax
\mciteBstWouldAddEndPuncttrue
\mciteSetBstMidEndSepPunct{\mcitedefaultmidpunct}
{\mcitedefaultendpunct}{\mcitedefaultseppunct}\relax
\EndOfBibitem
\bibitem[Giri and Schatz(2022)Giri, and Schatz]{Giri2022Manipulating}
Giri,~S.; Schatz,~G. Manipulating Two-Photon Absorption of Molecules through Efficient Optimization of Entangled Light. \emph{The journal of physical chemistry letters} \textbf{2022}, 10140--10146\relax
\mciteBstWouldAddEndPuncttrue
\mciteSetBstMidEndSepPunct{\mcitedefaultmidpunct}
{\mcitedefaultendpunct}{\mcitedefaultseppunct}\relax
\EndOfBibitem
\bibitem[Dayan \latin{et~al.}(2004)Dayan, Pe’er, Friesem, and Silberberg]{Dayan2004Two}
Dayan,~B.; Pe’er,~A.; Friesem,~A.; Silberberg,~Y. Two photon absorption and coherent control with broadband down-converted light. \emph{Physical review letters} \textbf{2004}, \emph{93 2}, 023005\relax
\mciteBstWouldAddEndPuncttrue
\mciteSetBstMidEndSepPunct{\mcitedefaultmidpunct}
{\mcitedefaultendpunct}{\mcitedefaultseppunct}\relax
\EndOfBibitem
\bibitem[Lu \latin{et~al.}(2022)Lu, Punj, and Orrit]{lu2022two}
Lu,~X.; Punj,~D.; Orrit,~M. Two-Photon-Excited Single-Molecule Fluorescence Enhanced by Gold Nanorod Dimers. \emph{Nano Letters} \textbf{2022}, \emph{22}, 4215--4222\relax
\mciteBstWouldAddEndPuncttrue
\mciteSetBstMidEndSepPunct{\mcitedefaultmidpunct}
{\mcitedefaultendpunct}{\mcitedefaultseppunct}\relax
\EndOfBibitem
\bibitem[Giannini \latin{et~al.}(2011)Giannini, Fernández-Domínguez, Heck, and Maier]{Giannini2011Plasmonic}
Giannini,~V.; Fernández-Domínguez,~A.; Heck,~S.; Maier,~S. Plasmonic nanoantennas: fundamentals and their use in controlling the radiative properties of nanoemitters. \emph{Chemical reviews} \textbf{2011}, \emph{111 6}, 3888--912\relax
\mciteBstWouldAddEndPuncttrue
\mciteSetBstMidEndSepPunct{\mcitedefaultmidpunct}
{\mcitedefaultendpunct}{\mcitedefaultseppunct}\relax
\EndOfBibitem
\bibitem[Akselrod \latin{et~al.}(2014)Akselrod, Argyropoulos, Hoang, Cirac{\`\i}, Fang, Huang, Smith, and Mikkelsen]{akselrod2014probing}
Akselrod,~G.~M.; Argyropoulos,~C.; Hoang,~T.~B.; Cirac{\`\i},~C.; Fang,~C.; Huang,~J.; Smith,~D.~R.; Mikkelsen,~M.~H. Probing the mechanisms of large Purcell enhancement in plasmonic nanoantennas. \emph{Nature Photonics} \textbf{2014}, \emph{8}, 835--840\relax
\mciteBstWouldAddEndPuncttrue
\mciteSetBstMidEndSepPunct{\mcitedefaultmidpunct}
{\mcitedefaultendpunct}{\mcitedefaultseppunct}\relax
\EndOfBibitem
\bibitem[Feng \latin{et~al.}(2015)Feng, You, Tian, Singamaneni, Liu, Duan, Lu, Xu, and Lin]{Feng2015Distance-Dependent}
Feng,~A.; You,~M.; Tian,~L.; Singamaneni,~S.; Liu,~M.; Duan,~Z.; Lu,~T.; Xu,~F.; Lin,~M. Distance-Dependent Plasmon-Enhanced Fluorescence of Upconversion Nanoparticles using Polyelectrolyte Multilayers as Tunable Spacers. \emph{Scientific Reports} \textbf{2015}, \emph{5}\relax
\mciteBstWouldAddEndPuncttrue
\mciteSetBstMidEndSepPunct{\mcitedefaultmidpunct}
{\mcitedefaultendpunct}{\mcitedefaultseppunct}\relax
\EndOfBibitem
\bibitem[Chen \latin{et~al.}(2012)Chen, Li, Yue, Xiao, and Gong]{chen2012plasmon}
Chen,~J.; Li,~Z.; Yue,~S.; Xiao,~J.; Gong,~Q. Plasmon-induced transparency in asymmetric {T}-shape single slit. \emph{Nano letters} \textbf{2012}, \emph{12}, 2494--2498\relax
\mciteBstWouldAddEndPuncttrue
\mciteSetBstMidEndSepPunct{\mcitedefaultmidpunct}
{\mcitedefaultendpunct}{\mcitedefaultseppunct}\relax
\EndOfBibitem
\bibitem[Izadshenas \latin{et~al.}(2018)Izadshenas, Zakery, and Vafapour]{izadshenas2018tunable}
Izadshenas,~S.; Zakery,~A.; Vafapour,~Z. Tunable slow light in graphene metamaterial in a broad terahertz range. \emph{Plasmonics} \textbf{2018}, \emph{13}, 63--70\relax
\mciteBstWouldAddEndPuncttrue
\mciteSetBstMidEndSepPunct{\mcitedefaultmidpunct}
{\mcitedefaultendpunct}{\mcitedefaultseppunct}\relax
\EndOfBibitem
\bibitem[Song \latin{et~al.}(2014)Song, Liu, Li, Song, Wei, and Song]{song2014dynamically}
Song,~J.; Liu,~J.; Li,~K.; Song,~Y.; Wei,~X.; Song,~G. Dynamically tunable plasmon-induced transparency in planar metamaterials. \emph{IEEE Photonics Technology Letters} \textbf{2014}, \emph{26}, 1104--1107\relax
\mciteBstWouldAddEndPuncttrue
\mciteSetBstMidEndSepPunct{\mcitedefaultmidpunct}
{\mcitedefaultendpunct}{\mcitedefaultseppunct}\relax
\EndOfBibitem
\bibitem[Luo \latin{et~al.}(2021)Luo, Wei, Lan, Wei, Meng, Liu, Yi, and Han]{luo2021dynamical}
Luo,~P.; Wei,~W.; Lan,~G.; Wei,~X.; Meng,~L.; Liu,~Y.; Yi,~J.; Han,~G. Dynamical manipulation of a dual-polarization plasmon-induced transparency employing an anisotropic graphene-black phosphorus heterostructure. \emph{Optics express} \textbf{2021}, \emph{29}, 29690--29703\relax
\mciteBstWouldAddEndPuncttrue
\mciteSetBstMidEndSepPunct{\mcitedefaultmidpunct}
{\mcitedefaultendpunct}{\mcitedefaultseppunct}\relax
\EndOfBibitem
\bibitem[Chu \latin{et~al.}(2009)Chu, Myroshnychenko, Chen, Deng, Mou, and de~Abajo]{Chu2009Probing}
Chu,~M.; Myroshnychenko,~V.; Chen,~C.-H.; Deng,~J.; Mou,~C.; de~Abajo,~F.~G. Probing bright and dark surface-plasmon modes in individual and coupled noble metal nanoparticles using an electron beam. \emph{Nano letters} \textbf{2009}, \emph{9 1}, 399--404\relax
\mciteBstWouldAddEndPuncttrue
\mciteSetBstMidEndSepPunct{\mcitedefaultmidpunct}
{\mcitedefaultendpunct}{\mcitedefaultseppunct}\relax
\EndOfBibitem
\bibitem[Tabakaev \latin{et~al.}(2021)Tabakaev, Montagnese, Haack, Bonacina, Wolf, Zbinden, and Thew]{tabakaev2021energy}
Tabakaev,~D.; Montagnese,~M.; Haack,~G.; Bonacina,~L.; Wolf,~J.-P.; Zbinden,~H.; Thew,~R. Energy-time-entangled two-photon molecular absorption. \emph{Physical Review A} \textbf{2021}, \emph{103}, 033701\relax
\mciteBstWouldAddEndPuncttrue
\mciteSetBstMidEndSepPunct{\mcitedefaultmidpunct}
{\mcitedefaultendpunct}{\mcitedefaultseppunct}\relax
\EndOfBibitem
\bibitem[Gorini \latin{et~al.}(1976)Gorini, Kossakowski, and Sudarshan]{gorini1976completely}
Gorini,~V.; Kossakowski,~A.; Sudarshan,~E. C.~G. Completely positive dynamical semigroups of N-level systems. \emph{Journal of Mathematical Physics} \textbf{1976}, \emph{17}, 821--825\relax
\mciteBstWouldAddEndPuncttrue
\mciteSetBstMidEndSepPunct{\mcitedefaultmidpunct}
{\mcitedefaultendpunct}{\mcitedefaultseppunct}\relax
\EndOfBibitem
\bibitem[Lindblad(1976)]{lindblad1976generators}
Lindblad,~G. On the generators of quantum dynamical semigroups. \emph{Communications in mathematical physics} \textbf{1976}, \emph{48}, 119--130\relax
\mciteBstWouldAddEndPuncttrue
\mciteSetBstMidEndSepPunct{\mcitedefaultmidpunct}
{\mcitedefaultendpunct}{\mcitedefaultseppunct}\relax
\EndOfBibitem
\bibitem[Novotny and Van~Hulst(2011)Novotny, and Van~Hulst]{novotny2011antennas}
Novotny,~L.; Van~Hulst,~N. Antennas for light. \emph{Nature photonics} \textbf{2011}, \emph{5}, 83--90\relax
\mciteBstWouldAddEndPuncttrue
\mciteSetBstMidEndSepPunct{\mcitedefaultmidpunct}
{\mcitedefaultendpunct}{\mcitedefaultseppunct}\relax
\EndOfBibitem
\bibitem[Scully and Zubairy(1997)Scully, and Zubairy]{scully1997quantum}
Scully,~M.; Zubairy,~M. \emph{Quantum Optics}; Quantum Optics; Cambridge University Press, 1997\relax
\mciteBstWouldAddEndPuncttrue
\mciteSetBstMidEndSepPunct{\mcitedefaultmidpunct}
{\mcitedefaultendpunct}{\mcitedefaultseppunct}\relax
\EndOfBibitem
\bibitem[AAT~Bioquest(2024)]{AATBioSpectrum[Atto700]}
AAT~Bioquest,~I. Spectrum [Atto 700]. \url{https://www.aatbio.com/fluorescence-excitation-emission-spectrum-graph-viewer/atto_700}, 2024; Accessed: 2024-05-31\relax
\mciteBstWouldAddEndPuncttrue
\mciteSetBstMidEndSepPunct{\mcitedefaultmidpunct}
{\mcitedefaultendpunct}{\mcitedefaultseppunct}\relax
\EndOfBibitem
\bibitem[AAT~Bioquest(2024)]{AATBioSpectrum[Atto610]}
AAT~Bioquest,~I. Spectrum [Atto 610]. \url{https://www.aatbio.com/fluorescence-excitation-emission-spectrum-graph-viewer/atto_610}, 2024; Accessed: 2024-05-24\relax
\mciteBstWouldAddEndPuncttrue
\mciteSetBstMidEndSepPunct{\mcitedefaultmidpunct}
{\mcitedefaultendpunct}{\mcitedefaultseppunct}\relax
\EndOfBibitem
\bibitem[AAT~Bioquest(2024)]{AATBioSpectrum[AttoRho6G]}
AAT~Bioquest,~I. Spectrum [Atto Rho6G]. \url{https://www.aatbio.com/fluorescence-excitation-emission-spectrum-graph-viewer/atto_rho6g}, 2024; Accessed: 2024-05-26\relax
\mciteBstWouldAddEndPuncttrue
\mciteSetBstMidEndSepPunct{\mcitedefaultmidpunct}
{\mcitedefaultendpunct}{\mcitedefaultseppunct}\relax
\EndOfBibitem
\bibitem[Izadshenas~Jahromi and S{\l}owik(2024)Izadshenas~Jahromi, and S{\l}owik]{izadshenas2024multiphoton}
Izadshenas~Jahromi,~S.; S{\l}owik,~K. Multiphoton absorption enhancement by graphene--gold nanostructure. \emph{Optics Letters} \textbf{2024}, \emph{49}, 3914--3917\relax
\mciteBstWouldAddEndPuncttrue
\mciteSetBstMidEndSepPunct{\mcitedefaultmidpunct}
{\mcitedefaultendpunct}{\mcitedefaultseppunct}\relax
\EndOfBibitem
\bibitem[Bharadwaj \latin{et~al.}(2009)Bharadwaj, Deutsch, and Novotny]{bharadwaj2009optical}
Bharadwaj,~P.; Deutsch,~B.; Novotny,~L. Optical antennas. \emph{Advances in Optics and Photonics} \textbf{2009}, \emph{1}, 438--483\relax
\mciteBstWouldAddEndPuncttrue
\mciteSetBstMidEndSepPunct{\mcitedefaultmidpunct}
{\mcitedefaultendpunct}{\mcitedefaultseppunct}\relax
\EndOfBibitem
\bibitem[Drobizhev \latin{et~al.}(2011)Drobizhev, Makarov, Tillo, Hughes, and Rebane]{drobizhev2011two}
Drobizhev,~M.; Makarov,~N.~S.; Tillo,~S.~E.; Hughes,~T.~E.; Rebane,~A. Two-photon absorption properties of fluorescent proteins. \emph{Nature methods} \textbf{2011}, \emph{8}, 393--399\relax
\mciteBstWouldAddEndPuncttrue
\mciteSetBstMidEndSepPunct{\mcitedefaultmidpunct}
{\mcitedefaultendpunct}{\mcitedefaultseppunct}\relax
\EndOfBibitem
\bibitem[Johansson \latin{et~al.}(2012)Johansson, Nation, and Nori]{johansson2012qutip}
Johansson,~J.~R.; Nation,~P.~D.; Nori,~F. QuTiP: An open-source Python framework for the dynamics of open quantum systems. \emph{Computer Physics Communications} \textbf{2012}, \emph{183}, 1760--1772\relax
\mciteBstWouldAddEndPuncttrue
\mciteSetBstMidEndSepPunct{\mcitedefaultmidpunct}
{\mcitedefaultendpunct}{\mcitedefaultseppunct}\relax
\EndOfBibitem
\bibitem[Johnson and Christy(1972)Johnson, and Christy]{johnson1972optical}
Johnson,~P.~B.; Christy,~R.-W. Optical constants of the noble metals. \emph{Physical Review B} \textbf{1972}, \emph{6}, 4370\relax
\mciteBstWouldAddEndPuncttrue
\mciteSetBstMidEndSepPunct{\mcitedefaultmidpunct}
{\mcitedefaultendpunct}{\mcitedefaultseppunct}\relax
\EndOfBibitem
\bibitem[Malitson(1965)]{malitson1965interspecimen}
Malitson,~I.~H. Interspecimen comparison of the refractive index of fused silica. \emph{Josa} \textbf{1965}, \emph{55}, 1205--1209\relax
\mciteBstWouldAddEndPuncttrue
\mciteSetBstMidEndSepPunct{\mcitedefaultmidpunct}
{\mcitedefaultendpunct}{\mcitedefaultseppunct}\relax
\EndOfBibitem
\bibitem[Jackson(2021)]{jackson}
Jackson,~J.~D. \emph{Classical electrodynamics}; John Wiley \& Sons, 2021\relax
\mciteBstWouldAddEndPuncttrue
\mciteSetBstMidEndSepPunct{\mcitedefaultmidpunct}
{\mcitedefaultendpunct}{\mcitedefaultseppunct}\relax
\EndOfBibitem
\bibitem[Izadshenas \latin{et~al.}(2023)Izadshenas, G{\l}adysz, and S{\l}owik]{izadshenas2023hybrid}
Izadshenas,~S.; G{\l}adysz,~P.; S{\l}owik,~K. Hybrid graphene-silver nanoantenna to control THz emission from polar quantum systems. \emph{Optics Express} \textbf{2023}, \emph{31}, 29037--29050\relax
\mciteBstWouldAddEndPuncttrue
\mciteSetBstMidEndSepPunct{\mcitedefaultmidpunct}
{\mcitedefaultendpunct}{\mcitedefaultseppunct}\relax
\EndOfBibitem
\bibitem[Izadshenas and S{\l}owik(2023)Izadshenas, and S{\l}owik]{izadshenas2023metasurface}
Izadshenas,~S.; S{\l}owik,~K. Metasurface for broadband coherent Raman signal enhancement beyond the single-molecule detection threshold. \emph{APL Materials} \textbf{2023}, \emph{11}, 081120\relax
\mciteBstWouldAddEndPuncttrue
\mciteSetBstMidEndSepPunct{\mcitedefaultmidpunct}
{\mcitedefaultendpunct}{\mcitedefaultseppunct}\relax
\EndOfBibitem
\end{mcitethebibliography}

\end{document}